\documentclass[12pt]{article}
\usepackage{graphicx}
\usepackage{amssymb}
\def\hybrid{\topmargin 0pt      \oddsidemargin 0pt
        \headheight 0pt \headsep 0pt
        \voffset=-0.5cm
        \textwidth 6.25in       % A4 paper
        \textheight 9.5in       % A4 paper
        \marginparwidth 0.0in
        \parskip 5pt plus 1pt   \jot = 1.5ex}
\catcode`\@=11
\def\marginnote#1{}

\newcount\hour
\newcount\minute
\newtoks\amorpm
\hour=\time\divide\hour by60 \minute=\time{\multiply\hour by60 \global\advance\minute by-\hour}
\edef\standardtime{{\ifnum\hour<12 \global\amorpm={am}%
        \else\global\amorpm={pm}\advance\hour by-12 \fi
        \ifnum\hour=0 \hour=12 \fi
        \number\hour:\ifnum\minute<10 0\fi\number\minute\the\amorpm}}
\edef\militarytime{\number\hour:\ifnum\minute<10 0\fi\number\minute}

\def\draftlabel#1{{\@bsphack\if@filesw {\let\thepage\relax
   \xdef\@gtempa{\write\@auxout{\string
      \newlabel{#1}{{\@currentlabel}{\thepage}}}}}\@gtempa
   \if@nobreak \ifvmode\nobreak\fi\fi\fi\@esphack}
        \gdef\@eqnlabel{#1}}
\def\@eqnlabel{}
\def\@vacuum{}
\def\draftmarginnote#1{\marginpar{\raggedright\scriptsize\tt#1}}
\def\draftlabel#1{{\@bsphack\if@filesw {\let\thepage\relax
   \xdef\@gtempa{\write\@auxout{\string
      \newlabel{#1}{{\@currentlabel}{\thepage}}}}}\@gtempa
   \if@nobreak \ifvmode\nobreak\fi\fi\fi\@esphack}
        \gdef\@eqnlabel{#1}}
\def\@eqnlabel{}
\def\@vacuum{}
\def\draftmarginnote#1{\marginpar{\raggedright\scriptsize\tt#1}}

\def\draft{\oddsidemargin -.5truein
        \def\@oddfoot{\sl preliminary draft \hfil
        \rm\thepage\hfil\sl\today\quad\militarytime}
        \let\@evenfoot\@oddfoot \overfullrule 3pt
        \let\label=\draftlabel
        \let\marginnote=\draftmarginnote
   \def\@eqnnum{(\theequation)\rlap{\kern\marginparsep\tt\@eqnlabel}%
\global\let\@eqnlabel\@vacuum}  }

\def\numberbysection{\@addtoreset{equation}{section}
        \def\theequation{\thesection.\arabic{equation}}}

\def\underline#1{\relax\ifmmode\@@underline#1\else
        $\@@underline{\hbox{#1}}$\relax\fi}

\def\titlepage{\@restonecolfalse\if@twocolumn\@restonecoltrue\onecolumn
     \else \newpage \fi \thispagestyle{empty}\c@page\z@
        \def\thefootnote{\fnsymbol{footnote}} }

\def\endtitlepage{\if@restonecol\twocolumn \else  \fi
        \def\thefootnote{\arabic{footnote}}
        \setcounter{footnote}{0}}  %\c@footnote\z@ }
\numberbysection \hybrid

\newcommand{\tr}{{\rm tr}}

\newcommand{\ti}[1]{\tilde{#1}}
\newcommand{\om}{\omega}

\newcommand{\al}{\alpha}

\newcommand{\be}{\beta}
\newcommand{\la}{\lambda}

\newcommand{\mat}[4]{\left(\begin{array}{cc}{#1}&{#2}\\{#3}&{#4}
\end{array}\right)}

\def\frak{\mathfrak}

\newtheorem{predl}{Proposition}

\newtheorem{theor}{Theorem}
\newtheorem{defi}{Definition}
\def\beq{\begin{equation}}
\def\eeq{\end{equation}}
\def\p{\partial}

\def\asl2{{\rm sl}(2, {\mathbb C})}
\def\GL2{{\rm GL}(2, {\mathbb C})}
\def\SL2{{\rm SL}(2, {\mathbb C})}

\begin{document}

\begin{titlepage}
\setcounter{page}{1}

\title{%On
Spectral Duality Between \\ Heisenberg Chain and Gaudin Model\\ $\ $}

\author{
 A.Mironov\thanks{Theory Department, Lebedev Physics Institute and ITEP, Moscow, Russia. $\ \ \ \ \ \ \ \ \ \ \ \ \ \ \ $ $\ \
\ \ \ \ \ \ \ $ E-mail: mironov@itep.ru; mironov@lpi.ru}\ \ \
 A.Morozov\thanks{ITEP, Moscow, Russia, E-mail:
morozov@itep.ru}\ \ \
 B.Runov\thanks{MIPT, Dolgoprudniy and ITEP, Moscow, Russia, E-mail: runovba@gmail.com}\ \ \
 Y.Zenkevich\thanks{Institute for Nuclear Research of the
Russian Academy of Sciences and ITEP, Moscow, Russia. E-mail: yegor.zenkevich@gmail.com}\ \ \
 A.Zotov\thanks{ITEP, Moscow, Russia. E-mail: zotov@itep.ru}
}

\date{}

\maketitle

\vspace{-7cm} \centerline{\hfill FIAN/TD-11/12}\centerline{ \hfill
ITEP-TH-20/12} \vspace{7cm}

\begin{abstract}
In our recent paper we described relationships between integrable systems inspired by the AGT conjecture. On the gauge theory side
an integrable spin chain naturally emerges while on the conformal field theory side one obtains
some special reduced Gaudin model. Two types of
integrable systems were shown to be related by the spectral duality. In this paper we extend the spectral duality to the case
of higher spin chains. It is proved that the $N$-site ${\rm GL}_k$ Heisenberg chain is dual to the  special reduced
$k+2$-points ${\rm gl}_N$ Gaudin model. Moreover, we construct an explicit Poisson map between the models at the classical level
by performing the Dirac reduction procedure and applying the AHH duality transformation.
\end{abstract}

\vfill\eject \tableofcontents

%\vfill

\end{titlepage}

\small{

\section{Introduction}
\label{sec:introduction}

This paper is a continuation of \cite{MMZZ} where the AGT correspondence  \cite{AGT} was studied at the level of integrable
systems \cite{SWint,NS,BS} (see also \cite{AGTN}-\cite{mty}). Two sides of the AGT relation
correspond to {\it a priori} different
types of integrable models which should actually coincide due to the AGT correspondence.
This leads to non-trivial predictions of equivalence of
different models and also illuminates what the equivalence exactly means. The full AGT correspondence associates the
conformal block of the Virasoro or $W$-algebra in two-dimensional conformal field theory with the LMNS integral \cite{LMNS}
(Nekrasov functions \cite{Nekr})) describing the two-parametric deformation of Seiberg-Witten theory by $\Omega$-background.
Classical integrable systems emerge when the both deformation parameters are brought to zero, while when only one of the
parameters goes to zero (the Nekrasov-Shatashvili limit \cite{NS}) the integrable system gets quantized \cite{BS}. We study
the correspondence between the AGT inspired integrable systems only in these two limiting cases.

In \cite{MMZZ} the simplest example of this kind was considered: the equivalence of the four-point conformal block and the
prepotential in the $SU(N)$ SUSY theory with $2N$ fundamental matter hypermultiplets and vanishing $\beta$-function. On the gauge
theory side the (classical) integrable system is known \cite{SWsc} to be the Heisenberg chain \cite{Heisen} which is
described by the spectral curve $\Gamma^{\hbox{\tiny{Heisen}}}(w,x):\det(w-T(x))=0$ with ${\rm GL}_2$-valued $N$-site
transfer-matrix $T(x)$ and the Seiberg-Witten \cite{SW} (SW) differential
$\hbox{d}S^{\hbox{\tiny{Heisen}}}(w,x)=x\frac{\hbox{d}w}{w}$. On the CFT side the corresponding integrable system was argued
to be some special reduced Gaudin model \cite{Gaudin1} defined by its spectral curve
$\Gamma^{\hbox{\tiny{Gaudin}}}(y,z):\det(y-L(z))=0$ with ${\rm gl}_N$-valued Lax matrix $L(z)$ and the SW differential
$\hbox{d}S^{\hbox{\tiny{Gaudin}}}(y,z)=y\hbox{d}z$.

{\em The spectral duality} \cite{AHH, Harnad2} generalizes the well-known self-duality of the Toda chain \cite{FT},
\cite{SWsc} and establishes relation between the two types of models in terms of the bispectral involution \cite{W1} which
interchanges the eigenvalue variable and the spectral parameter. In our case the spectral duality at the classical
level states that the change of variables $z=w$, $y=x/w$ relates the curves and the SW differentials of the two integrable
systems:
 \begin{equation}\label{q51007}
  \begin{array}{c}
  \Gamma^{\hbox{\tiny{Gaudin}}}(y,z)=\Gamma^{\hbox{\tiny{Heisen}}}(w,x)\,,
  \\
  \
  \\
\hbox{d}S^{\hbox{\tiny{Gaudin}}}(y,z)=\hbox{d}S^{\hbox{\tiny{Heisen}}}(w,x)\,.
  \end{array}
 \end{equation}
The quantum version of  the duality emerges from the exact quasi-classical quantization of the spectral curves based on the
corresponding SW differentials:
 \begin{equation}\label{qq10205}
 \begin{array}{c}
{\hat\Gamma}^{\hbox{\tiny{Heisen}}}(z,\hbar z\partial_z)\Psi^{\hbox{\tiny{Heisen}}}(z)=0\,,
 \end{array}
 \end{equation}
 \begin{equation}\label{qq10208}
 \begin{array}{c}
\hat\Gamma^{\hbox{\tiny{Gaudin}}}(\hbar
\partial_z,z)\Psi^{\hbox{\tiny{Gaudin}}}(z)=0
 \end{array}
 \end{equation}
with some choice of ordering. In \cite{MMZZ} the quantum spectral duality was proved in the form:
 \begin{equation}\label{qq102082}
 \begin{array}{c}
{\hat\Gamma}^{\hbox{\tiny{Heisen}}}(z,\hbar z\partial_z)\sim\hat\Gamma^{\hbox{\tiny{Gaudin}}}(\hbar
\partial_z,z).
  \end{array}
 \end{equation}
thus,
 \begin{equation}\label{qq10207}
 \begin{array}{c}
\Psi^{\hbox{\tiny{Heisen}}}(z)=\Psi^{\hbox{\tiny{Gaudin}}}(z)\,.
  \end{array}
 \end{equation}

{\bf The purpose of the paper} is twofold: first, to extend the results of \cite{MMZZ} to the case of higher spin chains. It
is proved that the $N$-site ${\rm GL}_k$ Heisenberg chain is dual to the  special reduced $k+2$-points ${\rm gl}_N$ Gaudin
model (Theorem 2, Section \ref{sec:classical-case}); and second, to construct an explicit Poisson map between the models, i.e.
to find an explicit change of variables at the classical level. For the second purpose, we perform the Poisson reduction
procedure via the Dirac brackets \cite{Dirac,BDOZ} in the Gaudin model and show that the reduced model exhibits the quadratic
Poisson structure (Propositions 1,2 in Section \ref{sec:gaudin-model}). Then, we apply the AHH duality transformation
\cite{AHH} and prove that the quadratic Poisson algebra of the dual reduced Gaudin model coincides with a natural quadratic
algebra underlying the Heisenberg spin chain (Theorem 3, Section \ref{sec:classical-case}). These results establish the exact
equivalence of the models:
 \begin{equation}\label{q1234}
 \begin{array}{|c|}
  \hline\\
\hbox{AHH}\Big\{zL^{\hbox{\tiny{Gaudin}}}(z)\Big\}(x)=T^{\hbox{\tiny{Heisen}}}(x){\prod\limits_j\frac{1}{x-x_j}}\,.\\ \ \\
\hline
  \end{array}
 \end{equation}
Here $L^{\hbox{\tiny{Gaudin}}}(z)$ is the Lax matrix of the Gaudin model and $T^{\hbox{\tiny{Heisen}}}(x)$ is the
transfer-matrix of the Heisenberg chain.

The paper is organized as follows: in Section 2 we review basic definitions and constructions related to the spectral dualities
in integrable systems. In Sections 3 and 4 the Gaudin model and the Heisenberg chain are described in detail. In Section 5
the classical spectral duality is proved and the explicit Poisson map between the models is presented. The quantum version of
the duality is described in Section 6. In the end we discuss some open problems and comment on relation of our results to the
previously known.

{\footnotesize{

\paragraph{Acknowledgments} The authors are grateful to A.Gorsky,  A.Zabrodin
and A.Zhedanov for useful comments and remarks. The work was partially supported by the Federal Agency for Science and
Innovations of Russian Federation under contract 14.740.11.0347 (A.Z., B.R. and Y.Z.), by NSh-3349.2012.2 (A.Mir., A.Mor. and
B.R.), by RFBR grants 10-02-00509 (A.Mir.), 10-02-00499 (A.Mor., and Y.Z.), 12-01-00482 (A.Z. and B.R.) and by joint grants
11-02-90453-Ukr, 12-02-91000-ANF, 12-02-92108-Yaf-a, 11-01-92612-Royal Society. The work of A.Zotov was also supported in
part by the Russian President fund MK-1646.2011.1.

}}

\section{Spectral Dualities and Integrable Systems}

\subsection{Spectral curves and Poisson structures}
Consider a classical algebraically integrable system \cite{Hit} with $g$ degrees of freedom. We assume that it is described
by the Lax matrix $L(z)\in{\mathrm{Mat}}(N)$ with the spectral parameter $z$, local coordinate on some complex curve
$\Sigma$. It means that the equations of motion with respect to any of the Poisson commuting Hamiltonians
$\{H_\al,H_\be\}=0$, $\al,\be=1...g$ (Liouville-Arnold integrability \cite{Ar}) have the Lax form:
 \begin{equation}\label{qq31}
  \partial_{t_\al}L(z)=\{H_\al,L(z)\}=[L(z),M_\al(z)].
 \end{equation}
The Lax equations may be interpreted as the compatibility condition of the {\em linear problem}:
 \begin{equation}\label{qq32}
  \left\{
  \begin{array}{l}
  (L(z)-\lambda)\phi=0,\\
  (\partial_{t_\al}+M_\al(z))\phi=0
  \end{array}
  \right.
 \end{equation}
The first equation gives rise to the {\em spectral curve} (a ramified $N$-fold covering of $\Sigma$)
 \begin{equation}\label{qq4}
  \Gamma(\lambda,z):\ \det(\lambda-L(z))=0.
 \end{equation}
which encodes the Hamiltonians and coupling constants\footnote{The coupling constants are the Casimir functions of the
corresponding Poisson brackets generated by some underlying classical $r$-matrix structure.} in the sense that $\tr\ L^k(z)$,
$1\leq k\leq N$ are generating functions of them. The spectral curve itself does not fix the integrable system. Indeed,
$\Gamma$ itself does not contain any information about the Poisson structure. Moreover, any Lax matrix $L(z)$ is defined up
to multiplication by an arbitrary function of $z$ and of integrals of motion. Therefore, some more ingredients should be added
to specify the integrable model. A straightforward way to do this is to introduce some classical $r$-matrix
\cite{FT,Skl4,STS,KRS} which
defines the Poisson brackets between any matrix elements of the Lax matrix. The basic examples of the $r$-matrix structures are
given by the linear
 \begin{equation}\label{qq401}
 \{L(z)\stackrel{\otimes}{_,}L(z')\}=[L(z)\otimes
    1+1\otimes L(z),r(z,z')]
 \end{equation}
and quadratic
 \begin{equation}\label{qq402}
  \{L(z)\stackrel{\otimes}{_,}L(z')\}=[L(z)\otimes L(z'),r(z,z')]
 \end{equation}
brackets. Each of these structures guarantees the involution $\{H_{i,k},H_{j,k}\}=0$ for the $z$-expansion coefficients
$H_{i,k}$ of $\tr\ L^k(z)$.

A general construction (see, e.g., \cite{Krich1}) provides solutions for the integrable model
%via the Abel map
in terms of the theta-functions on the Jacobian variety $\hbox{Jac}(\Gamma)$: the Liouville torus of the integrable
system\footnote{In concrete cases some additional factorization  may be required to make $\hbox{genus}(\Gamma)=g$.}. This
construction implies, of course, some Poisson (or symplectic) structure and some choice of Darboux coordinates on the phase
space. In fact, in \cite{Krich1,FT,Krich2} the canonical variables are chosen in accordance with the {\em separation of
variables} (SoV) recipe \cite{SoV,SoV2}. This recipe provides a set of variables with the Poisson brackets
 \begin{equation}\label{qq5}
 \{\lambda_\al,z_\be\}=h_\al(\lambda_\al,z_\al)\delta_{\al\be},\ \ \{\lambda_\al,\lambda_\be\}=\{z_\al,z_\be\}=0,\ \
 \al,\be=1...g,
  \end{equation}
where $h_\al$ are some functions. Each pair $(\lambda_\al,z_\al)$ lies on the spectral curve, i.e.
 \begin{equation}\label{qq6}
  \Gamma(\lambda_\al,z_\al)=0,\ \ \al=1...g.
 \end{equation}
{\bf Remark:} (\ref{qq6}) can be viewed as "separated equations" which appear in the SoV method when the generating
function of the canonical map is taken in the form ${\mathcal S}=\sum\limits_{\al=1}^g {\mathcal S}_\al(z_\al,c_1,...,c_g)$,
where $c_\al$ are fixed values of the Hamiltonians $H_\al$ that correspond to the given point of the moduli space of
(spectral) curves.
Then equations (\ref{qq6}) are equivalent to $\la_\al=\frac{\p
{\mathcal S}_\al}{\p z_\al}$. By the Liouville-Arnold definition, the integrable model is the Lagrangian bundle which base is the
set of values of the first integrals $H_\al$ and the fiber is the Lagrangian submanifold isomorphic to $\hbox{Jac}(\Gamma)$.
The existence of the separated variables means that the Lagrangian submanifold is globally presented as a product of $g$
copies $\Gamma\times...\times\Gamma$ of the spectral curve. The recipe \cite{SoV,SoV2} provides a concrete set of separated
variables. These are the poles $z_\al$ of the "properly normalized" Baker-Akhiezer function $\phi$ (\ref{qq32})  and the dual
variables (in the simplest case of the normalization) are the corresponding eigenvalues $\lambda_\al$ for (\ref{qq401}) or
$\log\lambda_\al$ for (\ref{qq402}). For example, in ${\rm sl}_2$ case the variables are defined as zeros of $L_{12}(z)$
while the dual ones are the values of $L_{11}(z_\al)$.

Therefore, the spectral curve is written in terms of separated variables. Then there reasonably exists on $\Gamma$
a generating differential $\hbox{d}S$ (of "$p\hbox{d}q$" type) which provides the action variables:
 \begin{equation}\label{qq7}
 \begin{array}{l}
  a_\al=\oint\limits_{A_\al}\hbox{d}S\,,
\\
 \frac{\!\partial}{\ \partial a_\al} \mathcal{F}_{\mathrm{SW}}=
  \oint\limits_{B_\al} \hbox{d}S,\ \  \al=1,...g\,.
  \end{array}
 \end{equation}
where $A_i, B_i$ are suitable cycles (homology basis) over  $\Gamma$ and $\mathcal{F}_{\mathrm{SW}}$ is the {\em
prepotential}. This differential (which is simply $\lambda\hbox{d}z$ for some cases) appears naturally in the context of
Seiberg-Witten theory \cite{SW} and is often called the {\em Seiberg-Witten (SW) differential}. Choosing a particular
$\hbox{d}S$ one fixes the functions $h_a$ in (\ref{qq5}). For example, if $\hbox{d}S=\lambda\hbox{d}z$ then $h_a=1$. The Poisson
structure (\ref{qq5}) can be also given in terms of the holomorphic symplectic form \cite{Krich2}:
 \begin{equation}\label{qq788}
\om=\delta(\hbox{d}S),
 \end{equation}
where $\delta$ denotes the exterior differential on the total space of the fibration of spectral curves and divisors
$[z_1,...,z_g]$.
%whose fiber is its Jacobian $\hbox{Jac}(\Gamma)$.

The Lax equations (\ref{qq31}) are invariant with respect to the gauge transformations
 \begin{equation}\label{qq8}
  L(z)\ \ \rightarrow\ g(z)L(z)g^{-1}(z),\ \ \ \ \
  M_\al(z)\ \ \rightarrow\ g(z)M_\al(z)g^{-1}(z)+\partial_{t_\al}g(z)g^{-1}(z)\,.
 \end{equation}
The corresponding $r$-matrix of course changes, while the spectral curve and the SW differential remain intact. In fact, we
deal with a special (still, a wide) class of Lax matrices having only simple poles on $\Sigma$. The gauge
transformations may  change residues and/or produce new poles. Then they connect different phase spaces equipped with
different Poisson (and $r$-matrix) structures \cite{LOZ1}.

Taking all the aforesaid into consideration, one may conclude that there are two ways to define the Poisson structure of an
integrable model. The first (gauge invariant) one is to choose the SW differential, the second (gauge dependent) way is to
define an $r$-matrix structure. We will use the first way and together with the spectral curve this defines the integrable system
in terms of the separated variables.

%The relationship between two approaches is described by the Separation of Variables recipe \cite{SoV,SoV2} or (equivalently)
%by the corresponding symplectic forms.

Let us briefly illustrate the construction for the SW theory which will be our primary focus in this paper. It is the
$\mathrm{SU}(N)$ gauge theory with $N_{\mathrm{f}}=2N$ matter hypermultiplets in the fundamental representation. The
corresponding classical integrable system is the $\rm{GL}_2$ Heisenberg XXX spin chain with $N$ cites \cite{APS,GGM} and the
spectral curve is written as follows:
\beq\label{eq:1}
 \Gamma^{\hbox{\tiny{Heisen}}}:  P(x) - \frac{w}{1+q} K_{+}(x) -
 \frac{q}{(1+q)w} K_{-}(x)=0\,,
\eeq
where
\begin{equation}
  \label{eq:202}
  P(x) = \prod_{k=1}^N (x-\phi_k)\,, \qquad K_{\pm} = \prod_{k=1}^N
  (x- m_k^{\pm})\,.
\end{equation}
The SW differential
\begin{equation}
  \label{qq801}
    \hbox{d}S=\frac{x\,dw}{w}\,.
\end{equation}
In the gauge theory the parameters $\phi_k$ are the (diagonal) vacuum expectation values of the adjoint scalar field
appearing in the $\mathcal{N} = 2$ vector multiplet ($\sum_k \phi_k = 0$), $m_k^{\pm}$ denote the masses of the
hypermultiplets.
%\footnote{One can see that the partition of the masses into ``$+$'' and ``$-$'' groups does not affect the prepotential.}.
The effective low-energy prepotential $\mathcal{F}_{\mathrm{SW}}$ is defined by the $A$- and $B$-periods
of the SW differential (\ref{qq801}) on the Riemann surface (\ref{eq:1}):
\begin{equation}\label{eq:3}
  a_i =  \oint\limits_{A_i} \frac{x\,dw}{w}\,,
\end{equation}
  \begin{equation}\label{eq:4}
  \frac{\partial \mathcal{F}_{\mathrm{SW}}}{\partial a_i} =
  \oint\limits_{B_i} \frac{x\,dw}{w}\,.
\end{equation}
{\bf Remark:} Notice that the SW differential can be also chosen as $\hbox{d}S=-\log(w)\hbox{d}x$. This differential gives
the same answer as (\ref{qq801}) for the integrals (\ref{qq7}) since both differentials correspond to the same representative
of the cohomology class $H^1(\Gamma,{\mathbb C})$.

\subsection{Quantization}

There are two natural ways to quantize the integrable model.  In accordance with its Poisson (symplectic) structure,
one can quantize either the $r$-matrix structure or the Poisson structure in separated variables (\ref{qq5}) corresponding to the
given SW differential. Let us start from the second possibility.

Considering the SW differential as a symplectic 1-form \cite{Krich2} on ${\mathbb C}^2$-plane $(y,z)$ yields a pair of
canonical variables $(p(y,z),q(z))$ which brings the SW differential to $\hbox{d}S(y,z)=p\hbox{d}q$. Then there is a natural
quantization of the spectral curve defined by the rule $(p,q)\rightarrow (\hbar\partial_q, q)$. Therefore, the quantization
follows from the  recipe based on the given SW differential in the quasiclassical form:
\begin{equation}
  \label{qq101}
  \hbox{with}\ \hbox{d}S={\lambda}\hbox{d}{z}\hbox{:}\ \  ({\lambda},{z})
  \ \stackrel{\hbox{\tiny{Quant}}}{\longrightarrow}\ (-i\hbar\partial_{{z}},{z}).
\end{equation}
with some choice of ordering. This choice may provide $\hbar$-corrections to the coefficients of the $\p_z$-expansion
of the quantum spectral curve. In the case of (\ref{qq801}), one has:
\begin{equation}
  \label{qq10}
  \Gamma(w,x)=0\ \stackrel{\hbox{\tiny{Quant}}}{\longrightarrow}\ \hat\Gamma=\Gamma(w,\hbar w\partial_w)\psi=0.
\end{equation}

The wave functions can be written in terms of the quantum deformation of the SW differential on the spectral curve, that is,
$\Psi(z)=\exp\left(-\frac{1}{\hbar}\int^q\hbox{d}S(\hbar)\right)$, where $\hbox{d}S(\hbar)=p(q,\hbar)\hbox{d}q$ and
$p(q,0)=p(q)|_\Gamma$. The monodromies of the wave function around $A$- and $B$- cycles of $\Gamma$ are given by the quantum
deformed action type variables \cite{BS}:
 \begin{equation}\label{qq10206}
 \begin{array}{l}
\Psi(z+A_i)=\exp\left(-\frac{1}{\hbar}a_i^\hbar\right)\Psi(z),\ \ a_i^\hbar=\oint\limits_{A_i}\hbox{d}S(\hbar)\,,
\\
  \
  \\
\Psi(z+B_i)=\exp\left(-\frac{1}{\hbar}\frac{\partial \mathcal{F}_{\hbox{\tiny{NS}}}}{\partial a_i^\hbar}\right)\Psi(z),\ \
\frac{\partial \mathcal{F}_{\hbox{\tiny{NS}}}}{\partial a_i^\hbar}=\oint\limits_{B_i}\hbox{d}S(\hbar)\,,
 \end{array}
 \end{equation}
where $\mathcal{F}_{\hbox{\tiny{NS}}}$ is the Nekrasov-Shatashvili limit \cite{NS} of the LMNS integral \cite{LMNS}. It
should be mentioned that we do not impose any boundary conditions which provide a valuable quantum problem, i.e. we do not
specify the wave functions explicitly. Instead, we analyze the differential operator of the quantum spectral curve.

The differential equation in the r.h.s. of (\ref{qq10}) is the {\em Baxter equation} \cite{Baxter}.  One also may choose another
quantization. For example, $(w,x)\rightarrow (\hbox{e}^{-\hbar\partial_x},x)$, i.e. $w$ maps to the shift operator. Then, the
Baxter equation is written in the difference form (Fourier dual). From the point of view of (\ref{qq101}), the latter case
corresponds to $\hbox{d}S=-\log(w)\hbox{d}x$. However, this differential gives the same answer as (\ref{qq801}) for
integrals (\ref{qq7}) since both differentials correspond to the same representative of the cohomology class
$H^1(\Gamma,{\mathbb C})$ as it was mentioned before.

Originally, the Baxter equation arises within the Quantum Inverse Scattering Method (QISM) \cite{Skl2,KRS}.
The QISM provides quantization
of the phase space and the corresponding Poisson structures (\ref{qq401})-(\ref{qq402}) via
\begin{equation}\begin{array}{c}
  \label{qq11}
  \hat{L}_1^\hbar(z)\hat{L}^\hbar_2(w)R^\hbar_{12}(z,w)=R^\hbar_{12}(z,w)\hat{L}^\hbar_2(w)\hat{L}^\hbar_1(z),\\
  R^\hbar_{12}(z,w)=1\otimes 1+\hbar\, r(z,w)+O(\hbar^2)
  \end{array}
\end{equation}
relations, where $L_1=L\otimes 1$, $L_2=1\otimes L$. Further development of the QISM requires some quantization conditions,
the {\em
Bethe equations}. In our approach we deal with exact quasi-classical equations and do not discuss Bethe-like equations
because we do not analyze concrete solutions.

Besides the approach proposed here, different quantizations of the Gaudin model are known (see, for example, \cite{FFR} and
\cite{Talalaev}). In \cite{Talalaev}  there was suggested a method for evaluation of $\det(\p_z+{\hat L}(z))$ as the generating
function of commuting quantum Hamiltonians. However, this method is based on the linear commutation relation in the
corresponding Lie algebra. In the classical case this corresponds to the linear Poisson-Lie structure. In our case (see
below) we deal with the reduced Gaudin model which is described by the quadratic brackets. Therefore, the method of
\cite{Talalaev} is non-applicable in our case (or, at least, requires some verification). In this paper we use
the recipe (\ref{qq101})
which provides the Baxter equation, i.e. a natural quantization of the spectral curve based on the separated variables.

\subsection{Bispectral problem and p-q duality} The notion of bispectral differential operators appeared
in the works of F.A.Gr\"{u}nbaum \cite{Grun} and J.J.Duistermaat \cite{Duis}. G.Wilson formulated it as {\em a bispectral
problem} \cite{W1}: construct the linear ordinary differential operator $\hat{L}=\sum_{j=0}^l L_j(z)\partial_z^j$ with
a nonempty family of eigenfunctions $\Psi(z,\lambda)$ depending smoothly on the spectral parameter $z$ such that
they are also
eigenfunctions of a linear ordinary differential operator $\hat{T}=\sum_{r=0}^m T_r(\lambda)\partial_\lambda^{\ r}$ with an
eigenvalue $g(z)$ which is a function of $z$:
 \begin{equation}\label{qq1}
 \begin{array}{c}
  \hat{L}(z,\partial_z)\Psi(z,\lambda)=f(\lambda)\Psi(z,\lambda),
  \\
 \hat{C}(\lambda,\partial_\lambda)\Psi(z,\lambda)=g(z)\Psi(z,\lambda)
 \end{array}
 \end{equation}
It appeared that for the Schr\"odinger operator $\hat{L}=\partial_z^2+V(z)$ the simplest solutions to the problem are given
by $V(z)=\frac{1}{z^2}$ (Bessel) and $V(z)=z$ (Airy) cases \cite{Duis}. Less trivial solutions can be obtained by applying
the rational Darboux transformations. They satisfy the KdV equation \cite{AMM}. After a link to the Calogero type systems was
also found \cite{Wilson,Kasman} it became  clear that the bispectral problem was closely related to the theory of integrable
systems \cite{bs0}. The bispectral problem resembles the quantum version of  {\em the p-q duality} \cite{MM5,p-q} while the case
of our interest is somewhat different.
%To be exact, the spectral duality is a part of the p-q duality?
Indeed, the p-q duality changes the coordinates of the model to the action variables of the dual model while the spectral
duality exchanges coordinates and momenta in separated variables. One can expect a certain relation between these two types
of dualities since the separation of variables  (\ref{qq5}) is "close" to the construction of the action-angle variables.
Indeed, after the variables are separated, the map to the action-angle variables is quite simple because it can be made
separately for each degree of freedom.
%Therefore, the p-q duality can be represented by steps...

At the same time, the p-q duality is very different from the spectral one. While the archetypal example of the spectral
self-duality is the Toda chain (see this example in Section 2.4), the p-q self-dual  model is the rational Calogero-Moser
system (and also the trigonometric Ruisenaars-Sneider and the hypothetic Double Elliptic Model \cite{MM5}).
Moreover, from the group theory interpretation of the p-q
duality it follows that the dual models possess Lax representations of the same size while in the spectral duality they are
different (say,  $2\times 2$ and $N\times N$).

\subsection{Spectral duality}
The duality we investigate in this paper is generated by {\em the bispectral involution} \cite{W1} which is simply a
change of arguments of the function
 \begin{equation}\label{qq2}
  b_{\hbox{\tiny{Wilson}}}:\ \Gamma(\lambda,z)\ \rightarrow\ \Gamma(z,\lambda)
 \end{equation}
corresponding to some spectral problem.
%\noindent {\em \underline{Definition:}}
 \begin{defi}Let a pair of (algebraically) integrable models be
 described by the
spectral curves $\Gamma(\lambda,z)=0$, $\Gamma'(\lambda',{z'})=0$ and the corresponding SW differentials
$\hbox{d}S(\lambda,z)$, ${\hbox{d}S}'(\lambda',{z'})$. Then the models are called spectrally dual at the classical level if
there exists a change of variables
$$
\la'=\la'(\la,z),\ \ z'=z'(\la,z)
$$
such that
 \begin{equation}\label{qq89}
  \Gamma(\lambda,z)=b_{\hbox{\tiny{Wilson}}}\left[\Gamma'\right](\la'(\la,z),z'(\la,z))=\Gamma'(z'(\la,z),\la'(\la,z))\,,
%  \equiv\check\Gamma(\check{z},\check\lambda)
 \end{equation}
and
  \begin{equation}\label{qq891}
  \hbox{d}S(\lambda,z)\cong \hbox{d}S(\la'(\la,z),z'(\la,z)),
 \end{equation}
where $\cong$ emphasizes that the SW differential for the integrable system is determined up to a full differential on the
spectral curve.
\end{defi}
 Let us give the very well-know

\noindent{\em \underline{Example {\bf\cite{FT}}:} The periodic Toda chain} can be described by both the ${\rm gl}(N)$-valued Lax
matrix
 \begin{equation}\label{qq100} { L}^{Toda}_{N\times N}(z) = \left(\begin{array}{ccccc}
p_1 & e^{{1\over 2}(q_2-q_1)} & 0 & & ze^{{1\over 2}(q_1-q_{N})}\\
e^{{1\over 2}(q_2-q_1)} & p_2 & e^{{1\over 2}(q_3 - q_2)} & \ldots & 0\\
0 & e^{{1\over 2}(q_3-q_2)} & p_3 & & 0 \\
 & & \ldots & & \\
\frac{1}{z}e^{{1\over 2}(q_1-q_{N})} & 0 & 0 & & p_{N}
\end{array} \right)
 \end{equation}
and the ${\rm GL}(2)$-valued  transfer-matrix \cite{FT}
 \begin{equation}\label{qq1010}
T^{Toda}_{2\times 2}(\lambda)=L_N(\lambda)...L_1(\lambda),\ \ \
L_i(\lambda) = \left(\begin{array}{cc} \lambda -p_i & e^{q_i} \\
-e^{-q_i} & 0
\end{array}\right), \ \ \ \ \ i = 1,\dots ,N
 \end{equation}
The spectral curves defined by these representations are related by
the bispectral involution, i.e.
 \begin{equation}\label{qq102}
\det(\lambda-L(z))=0\ \ \ \hbox{and}\ \ \ \ \det(z-T(\lambda))=0
 \end{equation}
coincide. The SW differential is the same in both cases
$\hbox{d}S=\lambda\frac{\hbox{d}z}{z}$. Therefore, {\em the periodic
Toda chain is a self-dual model.}

In quantum case we use the quantization scheme (\ref{qq7}) with some
choice of ordering.
 \begin{defi} Let two  integrable models be described by the Baxter equations
 \begin{equation}\label{qq103}
  \hat\Gamma\Psi=0\ \ \
  \hbox{and}\ \ \ \hat{\Gamma}'\Psi'=0\,.
 \end{equation}
They are called spectrally dual at the quantum level if their Baxter equations coincide.
 \end{defi}

In this paper we  prove that the special reduced ${\rm gl}_N$ Gaudin model is spectrally dual to the  XXX Heisenberg chain at
the classical and quantum levels. Moreover, we present an explicit Poisson map between the models at the classical level.

%\noindent{\em \underline{Statement [Theorems 1,2]:}
%First, let us discuss the relation of our results
%{\bf{Relations to the previously known constructions of dualities.}}

{\bf Remark:} At the classical level, the coincidence of the spectral curves was mentioned in \cite{Mironov:2010qe} for $N=2$.
For arbitrary $N$ the general form of the spectral curve for the Gaudin model was given in \cite{Yamada}. In quantum case the
Baxter equation for the ${\rm gl}_2$ Gaudin model was derived in \cite{MiTa}.

\section{Gaudin Model}\label{sec:gaudin-model}
%
%\subsubsection*{General Description}
%\label{sec:general-description-1}
%
Let $z$ be a local coordinate on ${\mathbb{CP}}^1$. The Lax matrix is a ${\rm gl}_N$-valued function $L^G(z)$ on
${{\mathbb{CP}}^1\backslash\{z_1,\dots,z_n\}}$ with only simple poles at $\{z_1,\dots,z_n\}$ and given residues
$\hbox{Res}_{z_c}L(z)=A^c\in{\rm gl}^*_N$:
\begin{equation}\label{q20}
  L^G(z)=\sum\limits_{c=1}^n\frac{A^c}{z-z_c}
\end{equation}
The spectral curve is
\begin{equation}\label{q2001}
  {\tilde\Gamma}^{{{Gaudin}}}(\tilde y,z):\ \ \ \ \det(\tilde y-L^G(z))=0.
\end{equation}

\subsection{Unreduced Gaudin model} The phase space of the Gaudin model
%\footnote{In many papers the unreduced Gaudin Model is commonly refered to as Gaudin Model.}
\cite{Gaudin1,Gaudin2} is a direct product of orbits of the coadjoint action of ${\rm GL}_N$:
\begin{equation}\label{q21}
  \begin{array}{c}{\tilde{\frak M}}^{{Gaudin}}={\mathcal O}_1\times\dots\times {\mathcal O}_n,\\
    \\ \dim {\tilde{\frak M}}^{{Gaudin}}=\sum\limits_{c=1}^n\dim{\mathcal O}_c.\end{array}
\end{equation}
This phase space is equipped with the Poisson-Lie brackets:
\begin{equation}\label{q2101}
  \{A^b_{ij},A^c_{kl}\}=\delta^{bc}\left(A^c_{kj}\delta_{il}-A^c_{il}\delta_{kj}\right)\,,\ \ b,c=1,...,n\,,
\end{equation}
which is generated by the linear $r$-matrix structure:
\begin{equation}\label{q2102}
  \begin{array}{c}\{L^G(z)\stackrel{\otimes}{,}L^G(w)\}=[L^G(z)\otimes
    1+1\otimes L^G(w),r(z,w)],\\ \\
    r(z,w)=\frac{1}{z-w}\sum\limits_{i,j=1}^N E_{ij}\otimes
    E_{ji}.\end{array}
\end{equation}
% \beq\label{q2101} \{A^b_\al,A^c_\be\}=-\delta^{bc}\sum\limits_\ga
% C_{\alpha\beta}^\gamma A^c_\gamma, \end{gather} where $A^c_\alpha$
% are coefficients in some basis $\{T_\alpha\}$:
% $A^c=\sum\limits_{\alpha} A^c_\alpha T_\alpha$ and $C_{\al\be}^\ga$
% are the structure constants of $\gln$ in this basis. The coordinates
% $\{A^c_\alpha\}$ on each orbit $A^c\in {\mathcal O}_c$ are chosen to
% be dual to the basis $\{T_\alpha\}$ of the Lie algebra ${\rm gl}(N,
% {\mathbb C})$.
The orbits are realized by fixation of the Casimir functions or eigenvalues of $\{A^c\}$, i.e.
\begin{equation}\label{q22}
  A^c=g A_0^c g^{-1}\,,\  g\in{\rm GL}_N\,,\ A_0^c=\hbox{diag}(\lambda_1,...,\lambda_N)\in{\rm gl}^*_N.
\end{equation}
The spectrum $A_0^c$ defines the dimension of ${\mathcal O}_c$. For example, in general case (when $\lambda_i$ are arbitrary)
\begin{equation}\label{q23}
  \dim {\mathcal O}^{\hbox{\tiny{ max}}}=N(N-1).
\end{equation}
In the case when $N-1$ eigenvalues coincide
\begin{equation}\label{q2301}
  \dim {\mathcal O}^{\hbox{\tiny{ min}}}=2(N-1).
\end{equation}
The later orbit can be parameterized in a ``quiver-like'' way \cite{KrB} using the vector (column) $\xi$ and the covector (row)
$\eta^T$:
\begin{equation}\label{q24}
  A=\xi\times \eta^T\,,\ \ A_{ij}=\xi_i\eta_j,\ \ \{\xi_i,\eta_j\}=\delta_{ij}
\end{equation} The
symmetry
 \begin{equation}\label{q2406}
\xi\rightarrow a\xi,\ \eta\rightarrow \frac{1}{a}\eta
\end{equation}
generates the ``conservation law'' $\sum\limits_{i=1}^N \eta_i\xi_i\!=\!\tr A\!=\!N\lambda\!=\!\hbox{const}$. Fixing the gauge
as $\eta_N=1$, one gets $\xi_N=N\lambda-\sum\limits_{i=1}^{N-1} \eta_i\xi_i$. After this reduction the Poisson brackets
between $\xi_i,\ \eta_j,\ i,j=1...N-1$ remain canonical.

\subsection{Specification of the model} One may also perform the reduction by the coadjoint action of ${\rm
  GL}_N$ acting on the ${\tilde{\frak M}}^{{Gaudin}}$
(\ref{q21}) as:
\begin{equation}\label{q25}
 L^{\hbox{\tiny{Gaudin}}}\rightarrow g L^{\hbox{\tiny{Gaudin}}}g^{-1}:\ \
 A^c\rightarrow g A^c g^{-1},\ \ \forall c,\ \ g\in{\rm GL}_N\,.
\end{equation}
It gives the first class constraint
\begin{equation}\label{q26}
  \sum\limits_{c=1}^n A^c=0
\end{equation}
and should be supplemented by some gauge fixation $\chi$. The reduced phase space is obtained by the Poisson reduction
\begin{equation}\label{q27}
  { {\frak M}}^{{Gaudin}}={\mathcal O}_1\times\dots\times {\mathcal O}_n//\hbox{Ad}_{{\rm GL}_N}\,.
\end{equation}
Since $\dim\hbox{Ad}_{{\rm GL}_N}=N^2-1$, one gets
\begin{equation}\label{q28}
  \dim { {\frak M}}^{{Gaudin}}=\sum\limits_{c=1}^n\dim{\mathcal O}_c-2(N^2-1).
\end{equation}
The SW differential is defined as
 \begin{equation}\label{q2802}
  \hbox{d}S^{\hbox{\tiny{Gaudin}}}=\tilde y\hbox{d}z
\end{equation}
This reduced model is of our main interest in this paper. Let us start with the example described in \cite{MMZZ}.

\subsubsection*{Gaudin Model on ${\mathbb{CP}}^1\backslash\{0,1,q,\infty\}$} \label{sec:specification-model}
Consider the case of four marked points $0, 1, q, \infty\footnote{${\rm SL}_2$ acts on
${{\mathbb {CP}}}^1$ by rational transformations, which allows one to fix any three of the marked point to be $0,\
  1,\ \infty$.}$ and let
$A^0$ and $A^\infty$ be generic orbits of the maximal dimension (\ref{q22}), (\ref{q23}), while $A^1$ and $A^q$ are those of
the minimal dimension (\ref{q2301})-(\ref{q24}), i.e.
\begin{equation}\label{q29}
  \begin{array}{c} A^1_{ij}=\xi_i^1\eta_j^1,\ \
    \{\xi^1_i,\eta^1_j\}=\delta_{ij},\ 1\leq i,j\leq N-1,\\ \\
    \eta_N^1=1,\ \ \xi_N^1=c_1-\sum\limits_{i=1}^{N-1}\eta^1_i\xi^1_i,\
    \ c_1=\tr A^1\end{array}
\end{equation}
and
\begin{equation}\label{q30}
  \begin{array}{c} A^q_{ij}=\xi_i^q\eta_j^q,\ \
    \{\xi^q_i,\eta^q_j\}=\delta_{ij},\ 1\leq i,j\leq N-1,\\ \\
    \eta_N^q=1,\ \ \xi_N^q=c_q-\sum\limits_{i=1}^{N-1}\eta^q_i\xi^q_i,\
    \ c_q=\tr A^q\end{array}
\end{equation}
Reduction by $\hbox{Ad}_{{\rm GL}_N}$ leads to (\ref{q26}):
\begin{equation}\label{q31}
  \varrho=A^0+A^1+A^q+A^\infty=0
\end{equation}
with the gauge fixed. The reduction procedure can be done in two steps which deal accordingly with the non-diagonal and
diagonal parts of the moment map (\ref{q31}).

\underline{The first part of the reduction:}
 \begin{equation}\label{q32}
 \begin{array}{l}
  \hbox{non-diag}(\varrho)=\hbox{non-diag}\left(A^0+A^1+A^q+A^\infty\right)=0\,,\\
  \chi=\hbox{non-diag}\left(A^\infty\right)=0\,.
  \end{array}
\end{equation}
i.e.
\begin{equation}\label{q33}
  \begin{array}{c}
    A^\infty=\Upsilon\equiv\hbox{diag}(\upsilon_1,...,\upsilon_N)\,.
    %\\ \\
    %A^0=-\Upsilon-A^1-A^q.
    \end{array}
\end{equation}

\underline{The second part of the reduction} involves the rest of the gauge group which is $\hbox{Stab}(\Upsilon)\simeq {\frak
H}$, the Cartan subgroup of ${\rm
  GL}_N$:
 \begin{equation}\label{q3208}
 \begin{array}{l}
  \hbox{diag}(\varrho)=\hbox{diag}\left(A^0+A^1+A^q+A^\infty\right)=0\,,\\
  \chi_{\frak H}=0\,,
  \end{array}
\end{equation}
where $\chi_{\frak H}$ are some fixing the $\hbox{Ad}_{\frak H}$ action.  It should be also mentioned here that the spectrum
of $A^0$ is fixed
\begin{equation}\label{q33003}
  A^0{\sim}\hbox{diag}(\mu_1,...,\mu_N),
\end{equation}
and therefore
\begin{equation}\label{q33004}
  \det(\kappa+\Upsilon+A^1+A^q)=\prod\limits_{i=1}^N(\kappa-\mu_i).
\end{equation}
Let us calculate the dimension of the reduced phase space. From (\ref{q28}) one has:
\begin{equation}\label{q3301}
  \begin{array}{c} \dim {
      {\frak M}}^{{Gaudin}}=\dim{\mathcal O}_0+\dim{\mathcal
      O}_1+\dim{\mathcal O}_q+\dim{\mathcal O}_\infty-2(N^2-1)\\ \\
    =2\dim {\mathcal O}^{\hbox{\tiny{ max}}}+2\dim {\mathcal
      O}^{\hbox{\tiny{
          min}}}-2(N^2-1)\stackrel{(\ref{q23}),(\ref{q2301})}{=}2N-2
  \end{array}
\end{equation}
After the first step (\ref{q33}) of the reduction (\ref{q27}) one obtains a $2\times 2(N-1)$-dimensional phase space. Then, the
second step of the reduction (by $\hbox{Ad}_{\hbox{Stab}(\Upsilon)}\simeq\hbox{Ad}_{\frak H}$, $\dim \hbox{Ad}_{\frak
H}=N-1$) leads to the dimension $2(N-1)$ as in (\ref{q3301}).

Below (in Section 5.3) we present the Poisson map of the Gaudin phase space to the Heisenberg chain phase space. We will
perform only the first step of the reduction and this gives us the exact coincidence of the Poisson structures under the
change of variables.

{\em Reflection symmetry.}
%\subsection{}
%\label{sec:s-duality}
%
One can easily see that our Gaudin model (\ref{q29})-(\ref{q33003}) possesses the following ${\mathbb Z}_2$ symmetry:
 \begin{equation}\label{eq:11}
  (q,z,A^0,A^1,A^q,\Upsilon)\ \longmapsto\ (q^{-1},z^{-1},\Upsilon,A^1,A^q,A^0)
\end{equation}
Indeed, the transformations (\ref{eq:11}) do not change the Lax matrix 1-from:
%(\ref{q3503}):
%
\begin{equation}\label{q4103}
\begin{array}{c}
 L^G(z)\hbox{d}z=  -\left(\Upsilon-\frac{A^1}{z-1}-q\frac{A^q}{z-q}\right)\frac{\hbox{d}z}{z}\ \
 \stackrel{(\ref{eq:11})}{\longmapsto}\ \  L^G(z)\hbox{d}z
\end{array}
\end{equation}
or
\begin{equation}\label{q4104}
\begin{array}{c}
 L^G(z)\ \ \stackrel{(\ref{eq:11})}{\longmapsto}\ \  -z^2 L^G(z)
\end{array}
\end{equation}
The described symmetry also remains unchanged in the quantum case since the quantization (\ref{q60}) $\ti y\mapsto \partial_z$ is
in agreement with (\ref{q4104}):
\begin{equation}\label{q4105}
\begin{array}{c}
 \partial_z-L^G(z)\ \ \stackrel{(\ref{eq:11})}{\longmapsto}\ \  -z^2 \left(\partial_z -L^G(z)\right).
\end{array}
\end{equation}

In the ${\rm gl}_2$ case, the reflection symmetry structure of this model written in the
elliptic parametrization \cite{Painl, ZZ}
was observed in \cite{LOZ2}.

\subsubsection*{Gaudin Model on ${{\mathbb{CP}}^1\backslash\{z_1,\dots,z_n\}}$}
In the general case, the construction is similar to the previous example. Let $z_1=0$ and $z_n=\infty$.
The specific configuration
of the Gaudin model under consideration is
 \begin{equation}\label{q3421}
  \begin{array}{l}
 \hbox{Spec}(A^0)=(\mu_1,...,\mu_N)\,,\\
 \hbox{Spec}(A^\infty)=(\upsilon_1,...,\upsilon_N)\,,\\
 A^c_{ij}=\xi^c_i\eta^c_j\,.
  \end{array}
 \end{equation}
The reduction constraints are
 \begin{equation}\label{q3422}
 \begin{array}{l}
  \hbox{non-diag}(\varrho)=0,\ \ \varrho=A^0+A^1+A^q+A^\infty\,,\\
  \chi=\hbox{non-diag}\left(A^\infty\right)=0\,
  \end{array}
\end{equation}
at the first step and
$$
 \begin{array}{l}
  \hbox{diag}(\varrho)=0,\ \ \chi_{\frak H}=0
  \end{array}
$$
at the second one. The dimension of the reduced phase space is equal to
 \begin{equation}\label{q3423}
  \begin{array}{c} \dim {
      {\frak M}}_1^{{Gaudin}}=(2N-2)(n-2)
  \end{array}
\end{equation}
after the first step of the reduction and finally
 \begin{equation}\label{q3424}
  \begin{array}{c} \dim {
      {\frak M}}^{{Gaudin}}=(2N-2)(n-3)\,.
  \end{array}
\end{equation}
To compute the number of parameters notice that whenever the number of the marked points is increased by one this adds two
constants (the coordinate of the point and the nontrivial eigenvalue of the minimal coadjoint orbit). Then, taking into account
(\ref{q42}) for the space of parameters ${\mathfrak{R}}$, one obtains
 \begin{equation}\label{q34241}
\dim \left({\mathfrak{R}}^{{Gaudin}}\right)=2N+1+2(n-4)=2(N+n)-7\,.
  \end{equation}
The Lax matrix
$$
L^{\hbox{\tiny{Gaudin}}}(z)=\frac{1}{z}A^0+\sum\limits_{c=2}^{n-1}\frac{1}{z-z_c}A^c
$$
under the constraint $\varrho=0$ can be written in the form
 \begin{equation}\label{q3426}
  \begin{array}{c}
L^{\hbox{\tiny{Gaudin}}}(z)=\frac{1}{z}\left(-\Upsilon+\sum\limits_{c=2}^{n-1}\frac{z_c}{z-z_c}A^c\right)
  \end{array}
 \end{equation}
since $A^0=-A^\infty-\sum\limits_{c=2}^{n-1}A^c$, $A^\infty=\Upsilon$.

\subsection{Poisson reduction}
The Dirac procedure \cite{Dirac} allows one to calculate the reduced Poisson structure in terms of the initial brackets on-shell.
Let us perform the reduction procedure.  There is no any distinguishable way to fix the action $\chi_{\frak H}$ (\ref{q3208}).
This is why we make only the first step (\ref{q3422}).

Suppose we deal with the constraints $h=(h_1,...,h_m)=0$ and the matrix of Poisson brackets between the constraints on-shell
 \begin{equation}\label{q3427}
  \begin{array}{c}
 C_{ij}=\left.\{h_i,h_j\}\right|_{h_i=0}
  \end{array}
 \end{equation}
is non-degenerate at the generic point of the phase space (the second class constrains by Dirac \cite{Dirac}). Then, the reduced
Poisson structure is given by the Dirac formula. For a pair of functions $f$ and $g$
 \begin{equation}\label{q3428}
  \begin{array}{c}
 \{f,g\}_D=\left.\left( \{f,g\}-\sum\limits_{i,j=1}^m \{f,h_i\}C^{-1}_{ij}\{h_j,g\}\right)\right|_{\hbox{\tiny{on-shell}}}
  \end{array}
 \end{equation}
In our case, there are $2(N^2-N)$ constraints
 \begin{equation}\label{q3429}
  \begin{array}{c}
 h=(\varrho_{ij},A^{\infty}_{ij})\,,\ \ i\neq j=1...N\,.
  \end{array}
 \end{equation}
The matrix $C$ (\ref{q3427}) is of the form:
 \begin{equation}\label{q3430}
  \begin{array}{c}
 C=\mat{\al+\be}{\be}{\be}{\be},\ \ \al=\{\sum\limits_{c=1}^{n-1}A^c,\sum\limits_{c=1}^{n-1}A^c\},\ \
 \be=\{A^{\infty},A^{\infty}\}\,.
  \end{array}
 \end{equation}
Therefore, the inverse matrix is equal to
 \begin{equation}\label{q3431}
  \begin{array}{c}
 C^{-1}=\mat{\al^{-1}}{-\al^{-1}}{-\al^{-1}}{\al^{-1}+\be^{-1}}\,.
  \end{array}
 \end{equation}
and it is not degenerate due to the arguments given in \cite{BDOZ}. A direct evaluation leads to the following results:
 \begin{predl}
 For the generic Gaudin model (\ref{q20}) the reduction corresponding to the first step  (\ref{q32}) gives the following
 reduced (Dirac) brackets
 \begin{equation}\label{q3432}
  \begin{array}{c}
  \{ A^a_{ij}, A^b_{kl} \}_D = \delta_{ab}(A^a_{il}\delta_{kj}-A^a_{kj}\delta_{il})-\\ \ \\
  - \sum\limits_{p \neq k} \frac{A^a_{ip}A^b_{p l}\delta_{jk}}
  {\varrho_{pp}-\varrho_{kk}-A^n_{pp}+A^n_{kk}} - \sum\limits_{p \neq i}
  \frac{A^b_{kp}A^a_{p j}\delta_{il}}
  {\varrho_{ii}-\varrho_{pp}+A^n_{pp}-A^n_{ii}}
  + \frac{A^a_{il}A^b_{kj} (1-\delta_{jl})} {\varrho_{ll}-\varrho_{jj}-A^n_{ll}+A^n_{jj}} +
  \frac{A^a_{kj}A^b_{il} (1-\delta_{ik})}
  {\varrho_{ii}-\varrho_{kk}-A^n_{ii}+A^n_{kk}}\,.
  \end{array}
 \end{equation}
 for $a,b\neq n$.
 \end{predl}

 \begin{predl}
For the special reduced Gaudin model (\ref{q3421}) the reduction corresponding to the first step  (\ref{q32}) gives the following
 reduced (Dirac) brackets
 \begin{equation}\label{q3433}
  \begin{array}{l}
  \{ \xi^a_i,\eta^b_j \}_D = -\delta_{ij}\left( \delta^{ab}+ \sum\limits_{p \neq
      i}\frac{\xi^a_p \eta^b_p}{\upsilon_{p}-\upsilon_{i} - \varrho_{pp} + \varrho_{ii}}
  \right)\,,\\
  \{ \xi^a_i,\xi^b_j \}_D =
  \frac{\xi^a_j\xi^b_i (1-\delta_{ij})}{\upsilon_j-\upsilon_i- \varrho_{jj} + \varrho_{ii}}\,,\\
  \{ \eta^a_i,\eta^b_j \}_D = \frac{\eta^a_j\eta^b_i(1-\delta_{ij})}{\upsilon_i-\upsilon_j- \varrho_{ii} + \varrho_{jj}}\,.
  \end{array}
 \end{equation}
 for $a,b\neq n$.
 \end{predl}
The formulae for the generic model (\ref{q3432}) are also valid for the special reduced model (\ref{q3421}).
Moreover, (\ref{q3432})
follows from (\ref{q3433}) via the initial parametrization $A^c_{ij}=\xi^c_i\eta^c_j$.

\subsection{Spectral curve}
Let us again  start from the example considered in \cite{MMZZ}.
 \begin{predl}{\bf \cite{MMZZ,Yamada}}
  The spectral curve for the Gaudin model defined by the data
  (\ref{q29})-(\ref{q33}) has the following form:
  \begin{equation}\label{q3501}
    \begin{array}{c} {\Gamma}^{\hbox{\tiny{Gaudin}}}( y,z): (1+q)\det(y\!+\!\Upsilon)\left(1\!+\!
        \frac{1}{1+q}\eta^1(y\!+\!\Upsilon)^{\!-1}\xi^1+\frac{q}{1+q}\eta^q(y\!+\!\Upsilon)^{\!-1}\xi^q\right) \\ \\
      =zP_{-\upsilon}(y)\!+z^{-1}qP_\mu(y),
    \end{array}
  \end{equation}
$$
P_\mu(y)\!=\!\prod\limits_{i=1}^N(y-\mu_i),\ \ P_{-\upsilon}\!=\!\prod\limits_{i=1}^N(y+\upsilon_i).
$$
 \end{predl}

\underline{\em{Proof}}:\vskip3mm

In order to compute the spectral curve (\ref{q2001}) we need the following simple Lemma:
%\paragraph{Lemma}

{\em  For any given invertible matrix $G\in\hbox{Mat}(N)$ and a pair of
  N-dimensional vectors $\xi$ and $\eta$:
  \begin{equation}\label{q34}
    \begin{array}{l} \hbox{1.}\ \
      \det(G+\xi\eta^T)=(1+\eta^T G^{-1}\xi)\det G,
    \end{array}
  \end{equation}
  \begin{equation}
    \label{q35}\begin{array}{l} \hbox{2.}\ \
      (G+\xi\eta^T)^{-1}=G^{-1}-\frac{1}{1+\eta^T
        G^{-1}\xi}G^{-1}\xi\eta^T G^{-1}.
    \end{array}
  \end{equation}
 }
Substituting (\ref{q33}) into the Lax matrix (\ref{q20}) with $A^c$ defined by (\ref{q29})-(\ref{q33004}), one gets:
\begin{equation}\label{q3503}
 L^{\hbox{\tiny{Gaudin}}}(z)=-\frac{1}{z}\left(\Upsilon-\frac{A^1}{z-1}-q\frac{A^q}{z-q}\right).
\end{equation}
Therefore, the spectral curve equation $\det (\ti{y}-L^G(z))=0$ can be written in the form:
\begin{equation}\label{q36}
  \det (\tilde y
  z+\Upsilon-\frac{1}{z-1}\xi^1\!\times\!\eta^1-\frac{q}{z-q}\xi^q\!\times\!\eta^q)=0
\end{equation}
Introduce
\begin{equation}\label{q38}
  y=\tilde y z.
\end{equation}
Applying (\ref{q34}), (\ref{q35}) from Lemma 1 twice, one obtains:
\begin{equation}\label{q37}
  \begin{array}{c}
    {\Gamma}^{\hbox{\tiny{Gaudin}}}( y,z):\ \ \ \
    \det(y+\Upsilon)\Big(1-\frac{1}{z-1}\eta^1(y+\Upsilon)^{-1}\xi^1-\frac{q}{z-q}\eta^q(y+\Upsilon)^{-1}\xi^q\\
    \\ +\frac{q}{(z-1)(z-q)}\Big(\eta^1(y\!+\!\Upsilon)^{\!-\!1}\xi^1 \cdot \eta^q(y\!+\!\Upsilon)^{\!-\!1}\xi^q\!-\!
    \eta^1(y+\Upsilon)^{\!-\!1}\xi^q \cdot
    \eta^q(y\!+\!\Upsilon)^{\!-\!1}\xi^1 \Big)\Big)\!=\! 0
  \end{array}
\end{equation}
The expression in the l.h.s. of (\ref{q37}) contains only simple poles at $z=1$ and $z=q$. The poles at $y=-\upsilon_i$ are
apparent. Indeed, it is easy to check that the second order poles are cancelled out in the second line of (\ref{q37}) while
the factor $\det(y+\Upsilon)$ cancell the first order poles. Moreover, let us compute $\det (y-A^0)$ for
(\ref{q33})-(\ref{q33004}) in the same way:
\begin{equation}\label{q39}
  \begin{array}{c} \det(y-A^0)\!=\!
    \det(y+\Upsilon)\Big(1+\eta^1(y+\Upsilon)^{-1}\xi^1+\eta^q(y+\Upsilon)^{-1}\xi^q\\
    \\ +\eta^1(y\!+\!\Upsilon)^{\!-\!1}\xi^1 \cdot \eta^q(y\!+\!\Upsilon)^{\!-\!1}\xi^q\!-\!
    \eta^1(y+\Upsilon)^{\!-\!1}\xi^q \cdot
    \eta^q(y\!+\!\Upsilon)^{\!-\!1}\xi^1\Big)=\prod\limits_{i=1}^N(y-\mu_i)
  \end{array}
\end{equation}
Then, the final answer is achieved by plugging the second line of (\ref{q39}) into the second line of (\ref{q37}):
\begin{equation}\label{q40}
  \begin{array}{c} z^2\det(y\!+\!\Upsilon)\!-\!z\det(y\!+\!\Upsilon)\left(
      \eta^1(y\!+\!\Upsilon)^{\!-1}\xi^1+q\eta^q(y\!+\!\Upsilon)^{\!-1}\xi^q \!+\!q\!+\!1\right) \\ \\ +q\det(y\!-\!A^0)\!=\!0,
  \end{array}
\end{equation}
$$
\det(y\!-\!A^0)\!=\!\prod\limits_{i=1}^N(y-\mu_i),\ \ \det(y\!+\!\Upsilon)\!=\!\prod\limits_{i=1}^N(y+\upsilon_i).
\hspace{2cm}\blacksquare
$$

Notice, that the case of the ${\rm sl}_N$ Gaudin model differs from the ${\rm gl}_N$ case by the shift $A^c\rightarrow
A^c-\frac{1}{N}\tr A^c$. Therefore, the spectral curves differ by
 \begin{equation}\label{q4002}
 y\rightarrow y+\frac{1}{N}\sum\limits_{z_c\in\{0,1,q\}} \frac{\tr A^c}{z-z_c}.
 \end{equation}

The space of parameters of the Gaudin spectral curve is described by the following set:
 \begin{equation}\label{q41}
  {\frak R}^{Gaudin}=\{\upsilon_1,...,\upsilon_N,\mu_1,...,\mu_N,\tr A^1,\tr A^q,q\}\,.
\end{equation}
Taking into account the possible shift of $y$, the number of independent parameters is equal to
\begin{equation}\label{q42}
  \dim {\frak R}^{Gaudin}=2N+1.
 \end{equation}
The case of arbitrary number of marked points (\ref{q3421}) is considered below (see Theorem 2).

\subsection{Simplest example: ${\rm gl}_2$ on ${\mathbb{CP}}^1\backslash\{0,1,q,\infty\}$} \label{sec:toy-exampl-painl}

Let us now calculate the spectral curve. In the ${\rm{gl}}_2$ case
 $$
 \det\sum\limits_i A_i=\sum\limits_i \det A_i+\sum\limits_{i<j}{\rm
   tr} A_i\tr A_j -\tr A_iA_j,\ \ A_i\in{\rm{gl}}_2.
 $$
The special case (\ref{q29})-(\ref{q33003}) for ${\rm{gl}}_2$ means that one deals with
 four $2\times 2$ matrices
$$A^0\sim\hbox{diag}(\mu_1,\mu_2),\ A^1=\xi^1\times \eta^1,\ A^q=\xi^q\times\eta^q,$$ where $\xi^1, \xi^q, \eta^1, \eta^q$ are
2-dimensional vectors and $A^\infty=\hbox{diag}(\upsilon_1,\upsilon_2)$ with the condition (\ref{q33}).  For the spectral
curve we have
 \begin{equation}\label{q43}
  \begin{array}{l}
    \tilde{y}^2-\left(\frac{\tr A^1}{z-1}+q\frac{\tr A^q}{z-q}+\frac{\mu_1+\mu_2}{z}\right)\tilde{y}+\\ \\
    +\frac{1}{z^2}\left(\det\Upsilon+q\frac{\tr A^1\tr A^q-\tr A^1A^q}{(z-1)(z-q)}
      -q\frac{\tr \Upsilon\tr A^q-\tr \Upsilon A^q}{z-q} -\frac{\tr \Upsilon\tr A^1-\tr \Upsilon A^1}{z-1}\right)=0,
  \end{array}
 \end{equation}
where in the second line we used $A^0=-\Upsilon-A^1-A^q$. Alternatively, one can make the shift (\ref{q4002})
$\tilde{y}\rightarrow \tilde{y}+\frac{1}{2}\left(\frac{\tr A^1}{z-1}+q\frac{\tr A^q}{z-q}+\frac{\mu_1+\mu_2}{z}\right)$,
which corresponds to the traceless case ${\rm{gl}}_2\rightarrow {\rm{sl}}_2$, i.e. $A_i\sim\hbox{diag}({\nu_i},{-\nu_i})$. In
this case the spectral curve can be written in the following form:
\begin{equation}\label{q44}
  \tilde{y}^2 - V(z)= -\frac{1-q}{z(z-1)(z-q)}H
\end{equation}
where $H$ is the Hamiltonian function on the phase space
\begin{equation}
  \label{eq:32}
  H = \tr \left[ A_q
  \left(A_0 + \frac{q}{q-1} A_1\right)\right] - \nu_q \left(\mu_1 + \mu_2 + \frac{2q}{q-1}
  \nu_1\right)
\end{equation}
and the potential reads
\begin{equation}\label{q45}
    V(z)=\frac{\nu_0^2}{z^2}+\frac{\nu_1^2}{(z-1)^2}+\frac{\nu_q^2}{(z-q)^2}-\frac{\nu_0^2+\nu_1^2+\nu_q^2-\nu_\infty^2}{z(z-1)}
\end{equation}
with $ \nu_0=\frac{\mu_1-\mu_2}{2},\ \ \nu_1=\frac{1}{2}\tr A^1,\ \ \nu_q=\frac{1}{2}\tr A^q,\ \
\nu_\infty=\frac{\upsilon_1-\upsilon_2}{2}$.

Remark: The coset space (\ref{q27}) in this case is the phase space of the Painlev{\'e} VI equation \cite{Painleve} in the
Schlesinger description \cite{Painleve2}.
%
%\section{spectral Duality}
%\label{sec:duality}
%

\subsection{AHH duality}

In \cite{AHH} M.R.Adams, J.Harnad and J.Hurtubise suggested a duality between the classical Gaudin-Schlesinger models of
different ranks and numbers of the marked points. Their description of the Gaudin model differs from ours by the constant
term $Y$ in the Lax matrix (or connection in the isomonodromic case):
 \begin{equation}\label{q71}
L^G_{AHH}(z)=Y+\sum\limits_{c=1}^M \frac{A^c}{z-z_c}\,,\ \ Y= \hbox{diag }(y_1,...,y_N)\,,\ \  A^c\in {\rm gl}_N\,.
 \end{equation}
The difference is essential, since $Y\neq 0$ leads to appearance of the second order pole at $\infty$ for
$L^G_{AHH}(z)\hbox{d}z$. The phase space is also different in this case.
It is a direct product of the coadjoint orbits (equipped with a
natural Poisson-Lie structure) factorized by  the stabilizer of $Y$: $\left({\mathcal O}^1\times\dots\times{\mathcal
O}^M\right)//\hbox{Stab}(Y)$.

In the case when all $A^c$ are of rank 1, the dual Lax matrix is the ${\rm gl}_M$-valued function with $\ti Y=\hbox{diag
}(z_1,...,z_M)$ and $N$ marked  points at $y_1,...,y_N$:
 \begin{equation}\label{q7101}
{\ti{L}}^G_{AHH}(z)=\ti Y+\sum\limits_{c=1}^N \frac{\ti A^c}{z-y_c}\,,\ \ \ti Y= \hbox{diag }(z_1,...,z_M)\,, \ \ \ti A^c\in
{\rm gl}_M\,.
 \end{equation}
The duality implies the relation between the spectral curves:
 \begin{equation}\label{q7102}
\det(\ti Y-z)\det(L^G_{AHH}(z)-\lambda)=\det(Y-\lambda)\det(\ti L^G_{AHH}(\lambda)-z)\,.
  \end{equation}
Dimensions of the phase spaces of the both models are equal to $2(N-1)(M-1)$ and the number of parameters is $2(N+M)-3$. Indeed, for
(\ref{q71}) dimension of the unreduced model is $\sum\limits_{c=1}^M \dim A^c=M\times 2(N-1)$. The reduction by the coadjoint
action of the Cartan subgroup $\hbox{Stab}(Y)\simeq {\frak H}\subset {\rm GL}_N$, $\dim \hbox{Ad}_{{\frak H}}=N-1$ leads to
$$
\sum\limits_{c=1}^M \dim A^c - 2\dim (\hbox{Stab}(Y))=M(2N-2)-2(N-1)=2(N-1)(M-1)\,.
$$

\section{Heisenberg Chain}
\label{sec:xxx-heisenberg-chain}
\subsection{${\rm GL}_2$ XXX Heisenberg chain}
%
%\subsubsection*{General Description} \label{sec:general-description}
%
Let $x$ be a local coordinate on ${\mathbb{CP}}^1$. Define the Lax operators as a set of ${\rm GL}_2$-valued functions
\begin{equation}\label{q1}
  L_i(x) = x - x_i + S^i, \qquad i = 1 \ldots N
\end{equation}
where $S^i=\sum\limits_{\alpha=1}^3 S^i_\alpha\sigma_\alpha$ are
matrices from ${\rm
  sl}_2$ and $\{ x_i\}$ is a set of points on
${\mathbb{CP}}^1$. Each $L_i(x)$ is assigned to the $i$-th site of
one-dimensional lattice. Then the transfer-matrix is defined as
\begin{equation}\label{q2}
  T(x)= V(q) L_N(x) \ldots L_1(x)
\end{equation}
where the "twist" $V(q)$ is a constant ${\rm GL}_2$-valued matrix. Following \cite{GGM} we choose $V$ depending on a
parameter $q$
\begin{equation}\label{q3}
  V=\left(\begin{array}{cc}{1}&{-\frac{q}{(1+q)^2}}\\{1}&{0}
    \end{array}\right),\ \ \ q=\hbox{const}
\end{equation}
with eigenvalues $\frac{q}{1+q}$ and $\frac{1}{1+q}$. The spectral curve and the SW differential are
\begin{equation}\label{q301}
  {\tilde\Gamma}^{\hbox{\tiny{Heisen}}}(\tilde w,x): \quad \det(\tilde w-T(x))=0,
\end{equation}
 \begin{equation}\label{q3011}
  \hbox{d}S^{{{XXX}}}=x\frac{\hbox{d}w}{w}\,.
\end{equation}
Expanding the determinant of the~$2 \times 2$ matrix, one obtains
\begin{equation}\label{q302}
  \tilde w-\tr  T(x) +{\tilde w}^{-1}\det T(x)=0\,.
\end{equation}
The Poisson structure consists of $N$ copies of the Lie--Poisson ${\rm
  sl}_2$-brackets at each site:
\begin{equation}\label{q4}
  \{ S^k_\alpha, S^l_\beta \} = \delta^{kl} \sqrt{-1} \, \varepsilon_{\alpha\beta\gamma}S^k_\gamma\,.
\end{equation}
These brackets are generated by the quadratic $r$-matrix structure:
\begin{equation}\label{q5}\begin{array}{c}
    \{L(z)\stackrel{\otimes}{,}L(w)\} = [r(z,w),L(z)\otimes
    L(w)]\,,\\ r(z,w) = \frac{1}{z-w} \left(1\otimes
      1+\sum\limits_{\alpha=1}^3\sigma_\alpha\otimes\sigma_\alpha\right)\,.\end{array}
\end{equation}
The values of the Casimir functions are defined by the eigenvalues $\hbox{Spec}(S^i)=(K_i,-K_i)$:
\begin{equation}\label{q6}
  K_i^2=\sum\limits_{\alpha=1}^3 S^i_\alpha S^i_\alpha\,, \qquad i=1
  \ldots N
\end{equation}
Thus, there is only a pair of independent variables at each site. Since $\det L_i(x)=(x-x_i)^2-K_i^2$ the spectral curve
(\ref{q302}) is now written in the following form
\begin{equation}\label{q7}
  \tilde w-\tr T(x) + {\tilde w}^{-1}\frac{q}{(1+q)^2}K^+(x)K^-(x)=0,
\end{equation}
\begin{equation}\label{q704}
  K^\pm(x) = \prod\limits_{i=1}^N(x-m_i^\pm)\,\ \ \ m_i^\pm=x_i\pm K_i.
\end{equation}
At this stage the phase space of the model is $2N$-dimensional
\begin{equation}\label{q8}
  {\tilde{\frak M}}^{\hbox{\tiny{Heisen}}}=\left\{S^i_\alpha,\ i=1 \ldots N,\
    \alpha=1 \ldots 3\ |\ K_i=\hbox{const}_i \in {\mathbb C} \,
  \right\}, \quad \dim {\tilde{\frak M}}^{\hbox{\tiny{Heisen}}}=2N
\end{equation}
Notice that the functions $K^\pm(x)$ are constant on ${\tilde{\frak
    M}}^{\hbox{\tiny{Heisen}}}$, while
\begin{equation}\label{q9}
  \tr T(x)=x^N + \sum\limits_{i=1}^N x^{i-1}H^{\hbox{\tiny{Heisen}}}_i \,,
\end{equation}
where $\left\{H^{\hbox{\tiny{Heisen}}}_i\right\}$ is a set of $N$ polynomial functions of degrees $N+1-i$ on ${\tilde{\frak
M}}^{\hbox{\tiny{Heisen}}}$ which mutually Poisson commute $\{H^{\hbox{\tiny{Heisen}}}_k, H^{\hbox{\tiny{Heisen}}}_j\} = 0$
with respect to (\ref{q4}) due to (\ref{q5}).

Let us also get in touch here with the parameters of the four dimensional
gauge theory: $m_i^{\pm}$ correspond to the masses of the $2N$ hypermultiplets in the fundamental representation, $q = e^{2\pi i
\tau}$ is related to the ultraviolet value of complex coupling constant $\tau$ and $x_i$ parameterize the vacuum expectation
values of the scalar fields.
\subsection{Reduced phase space and spectral curve} \label{sec:reduced-phase-space}
The Hamiltonian $H^{\hbox{\tiny{Heisen}}}_N$ is linear:
\begin{equation}\label{q10}
  H^{\hbox{\tiny{Heisen}}}_N=\tr\left(V\sum\limits_{i=1}^N S^i\right)-\sum\limits_{i=1}^N x_i.
\end{equation}
It means conservation of the projection of the total ``spin'' along the
``vector part'' of $V$: $\vec{V}=\frac{1}{2}\tr\left(
  V{\vec{\sigma}}\right)$. Obviously, it is generated by the adjoint
action of $g\in{\rm GL}_2$ which keeps $V$ unchanged, i.e.  $g\in\hbox{Stab}(V)$:
\begin{equation}\label{q11}
  g\in{\rm GL}_2:\ \ S^i\rightarrow g S^i g^{-1},\ \ V\rightarrow g V g^{-1}=V
\end{equation}

By analogy with the usage of the center of mass frame in many-body
systems (when the total momentum is conserved) we are going to resolve
the equation\footnote{The natural choice from the point of view of the gauge theory is to
set $H_N = 0$ because this eliminates the $U(1)$ factor and leaves the $SU(N)$ theory. However, we keep the constant
arbitrary.}
\begin{equation}\label{q12}
  H^{\hbox{\tiny{Heisen}}}_N=\hbox{const}
\end{equation} with
respect to any variable, say $S^N_\alpha$, and to reduce the phase space ${\tilde{\frak M}}^{\hbox{\tiny{Heisen}}}$ by 2
dimensions: the constraint (\ref{q12}) should be accompanied by some "gauge fixing" constraint, which corresponds to a
choice of nontrivial action (\ref{q11}) on $S^N$. Let us denote this second constraint as $\chi_N(S^N)=0$.

It is easy to show that the Poisson brackets (\ref{q4}) between $S^k,\ k=1...N-1$ are not changed by this reduction.
Finally, the reduced phase space
\begin{equation}\label{q13}
  \begin{array}{c}{ {\frak M}}^{\hbox{\tiny{Heisen}}}=\left\{S^i_\alpha,\
      i=1...N,\ \alpha=1...3\ |\ K_i=\hbox{const}_i,\
      H^{\hbox{\tiny{Heisen}}}_N=\hbox{const}, \chi_N(S^N)=0\right\} \\ \\ \dim {
      {\frak M}}^{\hbox{\tiny{Heisen}}}=2N-2
  \end{array}\end{equation}
has dimension $2N-2$ and the integrable dynamics is generated by $H^{\hbox{\tiny{Heisen}}}_1,\ldots
,H^{\hbox{\tiny{Heisen}}}_{N-1}$ with $S^N=S^N(S^1,\ldots ,S^{N-1})$ on-shell.

Let us also rewrite the spectral curve (\ref{q301}). Substituting $\tilde w=w\frac{1}{1+q}K^+(x)$ into (\ref{q302}), one gets
\begin{equation}\label{q14}
  \Gamma^{\hbox{\tiny{Heisen}}}(w,x):\ \ \tr
  T(x)-\frac{1}{1+q}wK^+(x)-\frac{q}{1+q}w^{-1}K^-(x)=0
\end{equation}
or
\begin{equation}\label{q15}
  \begin{array}{c}x^N\!+\!\frac{1}{1-q}\left(\frac{1}{w-1}\!-\!\frac{1}{w-q}\right)\left[w^2\Delta
      K^+(x)\!-\!w(1+q)\Delta\tr T(x)\!+\!q\Delta K^-(x)\right]\!=\!0,
  \end{array}
\end{equation}
where $\Delta f(x)=f(x)-x^N$. The ultimate curve depends on $N-1$
Hamiltonians and on the following set of $2N+2$ parameters:
\begin{equation}\label{q16}
  {\frak R}^{\hbox{\tiny{Heisen}}}=\{x_1,...,x_N,K_1,...,K_N, H_N,q\}\stackrel{(\ref{q704})}{\simeq}
  \{m_1^\pm,...,m_N^\pm, H_N,q\}.
\end{equation}
In fact, one of $\{x_i\}$ can be moved away by the shift of
$x$. Therefore, the number of independent parameters is equal to
\begin{equation}\label{q1601}
  \dim {\frak R}^{\hbox{\tiny{Heisen}}}=2N+1.
\end{equation}

% We do not include $H_N=\hbox{const}$ in the list because the
% corresponding term $x^{N-1}H_N$ can be removed by the shift
%
\subsection{Simplest example: 2-site chain}\label{sec:toy-example:-2}
Let
\begin{equation}\label{q1701}
  \tr \, T(x)=x^2+H_2 x+H_1,\ \ K^\pm=(x-m_1^\pm)(x-m_2^\pm)
\end{equation}
Substituting it into (\ref{q14}) with the change of variables $x=\tilde{y}z$, $w=z$,
one has
\begin{equation}\label{q1702}
  \tilde{y}^2z^2 + H_2 \tilde{y}z+H_1=\frac{z}{1+q}(\tilde{y}z-m_1^+)(\tilde{y}z-m_2^+)+
  \frac{1}{z}\frac{q}{1+q}(\tilde{y}z-m_1^-)(\tilde{y}z-m_2^-)
\end{equation}
This equation is easily reduced to the form
\begin{equation}\label{q1703}
  \tilde{y}^2-b(z)\tilde{y}+c(z)=0,
\end{equation}
with
\begin{equation}\label{q1704}
  \begin{array}{l}
    b(z)=\frac{(q+1)H_2+q(m_1^++m_2^+)+(m_1^-+m_2^-)}{(q-1)(z-q)} +\frac{(m_1^-+m_2^-)}{z}-
    \frac{(q+1)H_2+(m_1^++m_2^+)+q(m_1^-+m_2^-)}{(q-1)(z-1)},
    \\
    c(z)=\frac{q+1}{q}\frac {m^-_{{1}}m^-_{{2}}-H_{1}}{z}+\frac{m_1^-m_2^- + qm_1^+m_2^+ -(1+q)H_1}{q(q-1)(z-q)}-
    \frac{qm_1^-m_2^- + m_1^+m_2^+ -(1+q)H_1}{(q-1)(z-1)}+\frac{m_1^-m_2^-}{z^2}.
  \end{array}
\end{equation}
Making a shift $\tilde{y}\rightarrow \tilde{y}+\frac{1}{2}b(z)$, one
comes to $\tilde{y}^2+c(z)-\frac{1}{4}b^2(z)=0$
\begin{equation}\label{q170401}
  \tilde{y}^2+c(z)-\frac{1}{4}b^2(z)=0
\end{equation}
which is rewritten in the form
\begin{equation}\label{q1705}
  \tilde{y}^2-V(z)=-\frac{1-q}{z(z-1)(z-q)}H
\end{equation}
where $H$ is a Hamiltonian (linear in $H_1$) and
\begin{equation}\label{q1706}
  \begin{array}{l}
    V(z)=\frac{m_0^2}{z^2}+\frac{m_1^2}{(z-1)^2}+\frac{m_q^2}{(z-q)^2}-\frac{m_0^2+m_1^2+m_q^2-m_\infty^2}{z(z-1)}
  \end{array}
\end{equation}
with
\begin{equation}\label{q1707}
  \begin{array}{l}
    m_0=\frac{m^-_1-m^-_2}{2},\ \ \ m_\infty=\frac{m^+_2-m^+_1}{2},
    \\
    m_1=\frac{1}{2(q-1)}\left((q+1)H_2+m^+_1+m^+_2+q(m^-_1+m^-_2)\right),
    \\
    m_q=-\frac{1}{2(q-1)}\left((q+1)H_2+m^-_1+m^-_2+q(m^+_1+m^+_2)\right).
  \end{array}
\end{equation}
This should be compared with (\ref{q44})-(\ref{q45}). Note that $H_2$ is a constant (\ref{q12}). Therefore, $m_1$ and $m_q$
are constant as well since they are independent of $H_1$. The choice of signs in (\ref{q1707}) does not follow from
(\ref{q1706}). Our choice of signs will be justified in Section \ref{sec:quantum}.

\subsection{Higher rank chain}
Let us now consider the ${\rm GL}_k$ model
\begin{equation}
    L^i(x) = x-x_i+S^i\,,\ \ S^i\in{\rm gl}_k
\end{equation}
with the transfer-matrix
\begin{equation}
    T(x)=V(q)L_1(x) L_2(x) ... L_N(x),
\end{equation}
where $V(q)=\hbox{diag}(v_1,...,v_k)$ is a diagonal matrix (its entries will be determined later) and the Poisson brackets are
\begin{equation}\label{q6710}
    \{ L_n(x)\stackrel{\otimes}{_,}L_m(y) \}=\delta_{nm} [r(x,y),L_m(x) \otimes L_m(y)]\,.
\end{equation}
The $r$-matrix is defined as
\begin{equation}
  r(x,y)=\frac{P_{12}}{x-y}\,,\ P_{12}=\sum\limits_{a,b=1}^k E_{ab}\otimes E_{ba}\,.
\end{equation}
Then
\begin{equation}
  \{ S_{ab}^i,S_{cd}^j \} = \delta^{ij}(S_{ad}^i\delta_{cb} - S^i_{cb}\delta_{ad})\,.
\end{equation}
By virtue of the $r$-matrix relations (\ref{q6710}), one has the same
brackets for the transfer-matrix:
 \begin{equation}
  \{ T(x)\stackrel{\otimes}{_,}T(y) \}=[r(x,y),T(x)\otimes T(y)]\,.
 \end{equation}
Therefore,
 \begin{equation}\label{q6711}
 \begin{array}{c}
\{T_{ab}(z),T_{cd}(w)\}=\frac{1}{z-w}\left(T_{cb}(z)T_{ad}(w)-T_{ad}(z)T_{cb}(w)\right)\,,
\\ \ \\
\{T_{ab}(z),T_{cd}(z)\}=T_{ad}(z)\p_z T_{cb}(z)-T_{cb}(z)\p_zT_{ad}(z)\,.
 \end{array}
 \end{equation}
Let us now represent the transfer-matrix in the form of a sum:
\begin{equation}\label{q67111}
  T(x) = \prod_{j=1}^N (x - x_j)\, V(q) \prod_{i=1}^N \left(1 +
    \frac{S^i}{x - x_i}\right) =
  \prod_{j=1}^N(x-x_j) \left( V(q) + \sum_{i=1}^N
    \frac{\tilde{S}^i}{x-x_i} \right)\,.
 \end{equation}
The ${\rm gl}_k$-valued coefficients $\tilde{S}^i$ are the residues of the expression $T(x)/\prod\limits_j(x-x_j)$, i.e.
 \begin{equation}\label{q6712}
  \tilde{S}^i = \frac{T(x_i)}{\prod_{j\neq i}(x_i-x_j)} =\frac{1}{2\pi i}
  \oint_{\gamma_i} \frac{T(x)\, dx}{\prod_{j} (x-x_j)}\,,
 \end{equation}
where $\gamma_i$ is a small contour encircling $x_i$. Let us now calculate the brackets for the new variables $\tilde{S}^i$
in the spin chain.
 \begin{predl}
 The Poisson brackets for $\tilde{S}^i$ (\ref{q6712}) are of the following form:
 \begin{equation}\label{q6713}
  \{ \tilde{S}^i_{ab},\tilde{S}^j_{cd} \} = (1-\delta^{ij})\frac{\tilde{S}^i_{cb}\tilde{S}^j_{ad}
    -\tilde{S}^i_{ad}\tilde{S}^j_{cb}}{x_i-x_j}+\delta^{ij}\left(\tilde{S}^i_{ad}V_{cb} - \tilde{S}^i_{cb}V_{ad} + \sum\limits_{k\neq
    i}\frac{\tilde{S}^i_{ad}\tilde{S}^k_{cb}-\tilde{S}^i_{cb}\tilde{S}^k_{ad}}{x_i
    - x_k}\right)\,.
 \end{equation}
 \end{predl}
\underline{\em{Proof}}:\vskip3mm
%
%\noindent
For $i \neq j$ the brackets are easily obtained from the $r$-matrix relations for $T(x)$ (\ref{q6711}):
\begin{equation}
  \{ \tilde{S}^i_{ab},\tilde{S}^j_{cd} \} = \frac{\tilde{S}^i_{cb}\tilde{S}^j_{ad}
    -\tilde{S}^i_{ad}\tilde{S}^j_{cb}}{x_i-x_j} \quad \hbox{for }
  i\neq j
\end{equation}
For $i=j$ the arguments of the transfer-matrices coincide. One can use either local expansion of (\ref{q6711}) or integral
representation (\ref{q6712}) to overcome this problem:
\begin{equation}\label{q6132}
  \begin{array}{c}
  \{ \tilde{S}^i_{ab},\tilde{S}^i_{cd} \} = \oint_{\gamma_i}
  \oint_{\gamma_i'} dx\,dy \{ T_{ab}(x),
    T_{cd}(y) \}\prod\limits_{k,j}\frac{1}{x-x_j}\frac{1}{y-x_k}\\
  = \oint_{\gamma_i} \oint_{\gamma_i'} dx\,dy
  \frac{ T_{cb}(x) T_{ad}(y) - T_{ad}(x) T_{cb}(y) }{(x-y)}\prod\limits_{k,j}\frac{1}{x-x_j}\frac{1}{y-x_k}
  = \oint_{\gamma_i}  dx
  \frac{ T_{cb}(x) \tilde{S}^i_{ad} - T_{ad}(x) \tilde{S}^i_{cb} }{(x
    - x_i) \prod\limits_{j}
    (x-x_j)}\,.
    \end{array}
  \end{equation}
Now we shift the contour $\gamma_i$ to infinity so that it transforms to small contours encircling $x_j$ for $j \neq i$
and the contour around $x = \infty$, all of them going counterclockwise. The integral is computed by residues:
\begin{equation}\label{q6131}
  \begin{array}{c}
  \{ \tilde{S}^i_{ab},\tilde{S}^i_{cd} \} = - \sum\limits_{k\neq i}
  \oint_{\gamma_k} dx \frac{ T_{cb}(x) \tilde{S}^i_{ad} - T_{ad}(x)
    \tilde{S}^i_{cb} }{(x - x_i)\prod\limits_{j}
    (x-x_j)} - \\
  -\oint_{\gamma_\infty} dx \frac{ T_{cb}(x) \tilde{S}^i_{ad} -
    T_{ad}(x) \tilde{S}^i_{cb} }{(x - x_i)\prod\limits_{j} (x-x_j)} = - \sum\limits_{k\neq i}
  \frac{ \tilde{S}^k_{cb} \tilde{S}^i_{ad} - \tilde{S}^k_{ad}
    \tilde{S}^i_{cb} }{x_k - x_i} -
  V_{cb} \tilde{S}^i_{ad} - V_{ad} \tilde{S}^i_{cb}  = \\ \ \\
  = \tilde{S}^i_{ad}V_{cb} - \tilde{S}^i_{cb}V_{ad} + \sum\limits_{k\neq
    i}\frac{1}{x_i
    - x_k}({\tilde{S}^i_{ad}\tilde{S}^k_{cb}-\tilde{S}^i_{cb}\tilde{S}^k_{ad}})\hfill\blacksquare
    \end{array}
  \end{equation}
Though the formulae given above are valid for generic $S^i$, we consider the case when the site-variables are the coadjoint
orbits (of ${\rm GL}_k$) of {\em the minimal dimension} (see (\ref{q2301})) in this paper:
 \begin{equation}\label{q61321}
 \dim(S^i)=2(k-1)\,,\ \forall i=1...N\,.
  \end{equation}
Then the dimension of the phase space is equal to
 \begin{equation}\label{q6133}
\dim \left({\mathfrak{M}}^{\hbox{\tiny{Heisen}}}\right)=N(2k-2)-(2k-2)=2(k-1)(N-1)
  \end{equation}
where "$-(2k-2)$" comes from the reduction by the Cartan subgroup of ${\rm GL}_k$ (\ref{q13}). To compute the number of
parameters note that, with increasing the rank by unit, one adds two more constants (the component of $V$ and the value of the
Hamiltonian corresponding to the action of the Cartan subgroup of ${\rm GL}_k$). Then, taking into account (\ref{q1601}), one
obtains
 \begin{equation}\label{q6134}
\dim \left({\mathfrak{R}}^{\hbox{\tiny{Heisen}}}\right)=2N+1+2(k-2)=2(N+k)-3
  \end{equation}
for the dimension of the space of parameters.

\section{Classical Duality} \label{sec:classical-case}
Let us first recall the result of \cite{MMZZ}.
\subsection{Duality for ${\rm GL}_2$ chain}
As it follows from Proposition 3 (Section 3.4), the spectral curves of the special reduced Gaudin model
(\ref{q29})-(\ref{q33003}) and the XXX chain have the same forms (\ref{q14}) and (\ref{q3501}). Moreover, the dimensions of the
phase spaces (\ref{q13}), (\ref{q3301}) and the spaces of parameters (\ref{q16})-(\ref{q1601}), (\ref{q41})-(\ref{q42}) can
be identified as well. This is summarized in the following

\begin{theor}
  The N-site  ${\rm GL}_2$ Heisenberg XXX chain defined by  (\ref{q1})-(\ref{q1601}) and  the
  ${\rm gl}_N$ Gaudin model
  (\ref{q25})-(\ref{q42}) are spectrally dual at the classical level with the change of variables:
  \begin{equation}\label{q51}
    \ \left\{\begin{array}{l} z=w\,,\\ \\
        \tilde y=\frac{x}{w}\,,
      \end{array}\right.
  \end{equation}
  the following identification of the parameters (\ref{q16}) and
  (\ref{q41}):
  \begin{equation}\label{q52}
    \left\{\begin{array}{l}
        m^+_i=-\upsilon_i,\ \ 1\leq
        i\leq N,\\ \\
        m^-_i=\mu_i,\ \ 1\leq
        i\leq N,\\ \\
        H^{\hbox{\tiny{XXX}}}_N=\frac{1}{1+q}\tr A^1+\frac{q}{1+q}\tr
        A^q+\sum\limits_{k=1}^N\upsilon_k
      \end{array}\right.
  \end{equation}
  and the following relation between the generating functions of the
  Hamiltonians:
  \begin{equation}\label{q53}
    \begin{array}{c}
      \tr T^{\hbox{\tiny{XXX}}}(y)= \det(y\!+\!\Upsilon)\left(1\!+\!
        \frac{1}{1+q}\eta^1(y\!+\!\Upsilon)^{\!-1}\xi^1+\frac{q}{1+q}\eta^q(y\!+\!\Upsilon)^{\!-1}\xi^q\right)
    \end{array}
  \end{equation}
\end{theor}

The simplest example of the duality comes from comparison of  (\ref{q1705})-(\ref{q1707}) with (\ref{q44})-(\ref{q45}).

\subsection{Duality for higher rank chains}\label{sec:higher-ranks}
All the statements of the previous sections work not only for the ${\rm GL}_2$ spin chain and the $4$-point Gaudin system,
but also for the ${\rm GL}_k$ spin chain and the $k+2$-point Gaudin model.
%Naturally, one should ensure that all the matrices $\tilde{S}^i$ and $k$ of the $k+2$ matrices $A^a$ are of the minimal rank.
%This ensures that these matrices can indeed be decomposed as in (\ref{eq:5}).
Indeed, from (\ref{q3424}), (\ref{q34241}) and (\ref{q6133}), (\ref{q6134}) one concludes that the dimensions of the phase
spaces and spaces of parameters are equal for the Gaudin model (\ref{q3421})-(\ref{q34241}) and the Heisenberg chain
(\ref{q6133})-(\ref{q6133}) if $n=k+2$.
 \begin{theor}
  The N-site  ${\rm GL}_k$ Heisenberg XXX chain (\ref{q67111})-(\ref{q6134}) and  the
  ${\rm gl}_N$ Gaudin model on ${{\mathbb{CP}}^1\backslash\{z_1,\dots,z_n\}}$
  (\ref{q3421})-(\ref{q34241}) are spectrally dual at the classical level with

  ${\bf 1.}$ the change of the spectral problem variables:
  \begin{equation}\label{q6211}
    \begin{array}{l} z=w\,,\\ \\
        \tilde y=\frac{x}{w}\,,
      \end{array}
  \end{equation}

 ${\bf 2.}$ the following identification of the parameters:
   \begin{equation}\label{q6210}
n=k+2\,,
  \end{equation}
  \begin{equation}\label{q6212}
    \begin{array}{l}
       V=\hbox{diag}(z_2,...,z_{n-1})\in{\rm gl}_k\,,
      \end{array}
  \end{equation}
      \begin{equation}\label{q6214}
    \begin{array}{l}
       A^\infty_{ii}\equiv\Upsilon_{ii}\equiv\upsilon_i=-x_i\,.
      \end{array}
  \end{equation}
 \end{theor}

 ${\bf 3.}$ and the following change of dynamical variables:
 \begin{equation}\label{q6213}
    \begin{array}{l}
       \left(V^{-1}\tilde{S}^i\right)_{ab}=\xi^a_i\eta^b_i\,,\ \ a,b=1...k\,,\ \  i=1...N\,,
      \end{array}
  \end{equation}

\underline{\em{Proof}}:\vskip3mm

Let us see explicitly how the identification between the two systems manifests itself at the level of spectral curves. One
writes for the spectral curve of the spin chain:
\begin{equation}
  \label{eq:8}
  \det_{k \times k} \left( z - \prod_{j=1}^N (x - x_j) V \left( 1 +
    \sum_{i=1}^N \frac{\xi_i \eta_i^T }{x - x_i} \right) \right)=0\,.
\end{equation}
For the $k+2$ point Gaudin system:
\begin{equation}
  \label{eq:6}
  \det_{N\times N} (yz - L_G) = \det_{N \times N} \left( yz + A_{\infty} - \sum_{a=1}^k \frac{z_a \xi^a
    (\eta^a)^T}{z - z_a} \right) = 0\,.
\end{equation}
We denote $yz$ as $x$ and assume as in the preceding sections that the matrix $A_{\infty}$ is diagonal with the diagonal entries
$\upsilon_i$, $i = 1\ldots N$. Then one has:
  \begin{equation}\label{eq:7}
    \begin{array}{c}
  \det_{N \times N} \left( x + A_{\infty} - \sum_{a=1}^k \frac{z_a \xi^a
      (\eta^a)^T}{z - z_a} \right) =\\
  = \prod\limits_{i=1}^N (x + \Lambda_i) \det_{N
  \times N} \left( 1 - (x + A_{\infty})^{-1} \sum_{a=1}^k \frac{z_a \xi^a
    (\eta^a)^T}{z - z_a} \right)\,.
    \end{array}
  \end{equation}
Let us introduce  $D = (x + A_{\infty})^{-1} \sum_{a=1}^k \frac{z_a \xi^a
  (\eta^a)^T}{z - z_a}$. Expanding the determinant into powers of $D$, one gets the
sum of terms, each term being a product of traces of $D^n$. A typical contribution looks like $c_{n_1,\ldots ,n_j} \tr
D^{n_1} \cdots D^{n_j}$. Let us point out that the coefficients $c_{n_1,\ldots ,n_j}$ do not depend on $N$, the size of the
matrix $L_G$. We are going to prove therefore that each trace of $D^n$ can be rewritten as the trace of the $n$-th power of a
$k \times k$ matrix. This will prove the equivalence between the two spectral curves.

The derivation goes as follows. Consider
  \begin{equation}\label{eq:903}
    \begin{array}{c}
  \tr_{N \times N} D^M = \sum\limits_{i_1\ldots i_M = 1}^N \sum\limits_{a_1\ldots
    a_M = 1}^k \frac{1}{x + \Lambda_{i_1}} \cdot \frac{z_{a_1} \xi_{a_1}^{i_1}
    \eta_{a_1}^{i_2}}{(z - z_{a_1})} \cdots \frac{1}{x +
    \Lambda_{i_M}} \cdot \frac{z_{a_M} \xi_{a_M}^{i_M}
    \eta_{a_M}^{i_1}}{(z - z_{a_M})}=\\
  = \sum\limits_{i_1\ldots i_M = 1}^N \tr_{k \times k} \left( (z - Z)^{-1} Z
    \frac{ \eta^{i_1} (\xi^{i_1})^T}{(x + \Lambda_{i_1})} \cdots
     (z - Z)^{-1} Z \frac{
       \eta^{i_M} (\xi^{i_M})^T}{(x + \Lambda_{i_M})} \right)=\\
   = \tr_{k \times k} \left[ \sum\limits_{i=1}^N \frac{(z - Z)^{-1} Z \eta^i (\xi^i)^T }{x+\Lambda_i} \right]^M\,,
    \end{array}
  \end{equation}
where the traces in the second and third lines are taken over the indices $a_i$ and $Z$ is the ${\rm gl}_k$-valued diagonal
matrix with the diagonal entries $z_a$:
$$
Z=\hbox{diag}(z_1,...,z_k)\,.
$$
Thus, one has:
  \begin{equation}\label{eq:10}
    \begin{array}{c}
  \det_{N \times N} \left( x\!+\! A_{\infty}\! -\! \sum_{a=1}^k \frac{z_a
      \xi^a
      (\eta^a)^T}{z - z_a} \right)
  = \prod\limits_{i=1}^N (x\! +\! \Lambda_i) \det_{k \times k} \left[ 1\! -\!
    \sum_{i=1}^N \frac{(z - Z)^{-1} Z \eta^i
      (\xi^i)^T }{x+\Lambda_i} \right]=\\
  = \frac{\prod\limits_{i=1}^N (x + \Lambda_i)}{\prod\limits_{a=1}^k (z - z_a)}
  \det_{k \times k} \left[ z - Z \left( 1 + \sum_{i=1}^N \frac{ \eta^i
        (\xi^i)^T }{x+\Lambda_i} \right) \right]\,.
    \end{array}
  \end{equation}
The spectral curves coincide provided one identifies $Z = V$\footnote{From now on, we use
the different overall normalization for $z$ and, hence, for $V$: that is,
for $k = 2$ we set $V =
  \mathrm{diag}\, \left(\frac{1}{1+q}, \frac{q}{1+q}\right)$ instead
  of $V = \mathrm{diag}\, (1, q)$.}, $\tilde{S}^i = V \xi^i
(\eta^i)^T$ and $\upsilon_i = - x_i$.\hfil$\blacksquare$

Remark: in fact, any Gaudin system can be rewritten as a spin chain by merging together several points containing minimal
rank orbits. This will correspond to merging of the eigenvalues of the matrix $V$ in the spin chain. Conversely, any spin
chain can be rewritten as a Gaudin system by analogously merging several of the points $x_i$. The matrix $A^{\infty}$ in the
Gaudin system then has several coinciding eigenvalues.

%Summing up, all the story very much resembles the AHH duality, the only nontrivial step being the rewriting of the transfer
%matrix for the spin chain as a sum of residues.

\subsection{Poisson map}
In this paragraph we show that the identification (\ref{q6213}) provides a Poisson map between the two models. Our strategy
is the following: we start from the Gaudin model (\ref{q3421})-(\ref{q3426}) with the quadratic Poisson brackets
(\ref{q3433}). Then we apply the AHH duality transformation (\ref{q7101}) to $zL^{\hbox{\tiny{Gaudin}}}$.
%, i.e. we exclude the common factor $\frac{1}{z}$.
After that we show that the ${\rm gl}_{n\!-2}$-valued residues of the dual model obey the same
Poisson structure as $\tilde{S}^i$ (\ref{q6713}).

Denote
\begin{equation}\label{q6151}
    B^i_{ab} = \xi^a_{i}\eta^b_{i}\in {\rm gl}_k\,.
\end{equation}
In fact, the proof of Theorem 2 follows by applying the AHH duality (\ref{q7102}). One can define the AHH dual of the
Gaudin model (\ref{q3426})
$$
L^{\hbox{\tiny{Gaudin}}}(z)=\frac{1}{z}\left(-\Upsilon+\sum\limits_{c=2}^{n-1}\frac{z_c}{z-z_c}A^c\right),\ \
\Upsilon=\hbox{diag}(\upsilon_1,...,\upsilon_N)
$$
following the recipe (\ref{q7101}):
\begin{equation}\label{q6152}
  \begin{array}{c}
 \hbox{AHH}\Big\{zL^{\hbox{\tiny{Gaudin}}}(z)\Big\}(x)=Z+\sum\limits_{i=1}^N\frac{ZB}{x+\upsilon_i}\,.
    \end{array}
  \end{equation}
However, the Poisson structures of both models are quadratic in our case (instead of the linear brackets in AHH \cite{AHH}).

\begin{theor}
Under the AHH duality transformation the phase spaces of the ${\rm gl}_N$ reduced $k+2$-point Gaudin model (\ref{q3421}),
(\ref{q3423}), (\ref{q3433}) and the XXX  Heisenberg chain (\ref{q67111})-(\ref{q6713}) are related by  (\ref{q6213})
$$
       \left(V^{-1}\tilde{S}^i\right)_{ab}=B^i_{ab}\,,\ \ a,b=1...k\,,\ \  i=1...N\,.
$$
This map is Poisson, i.e. the Poisson brackets for $B^i$ (\ref{q6151}) generated by the Dirac brackets (\ref{q3433}) are
of the same form as for $V^{-1}\tilde{S}^i$ (\ref{q6713}):
 \begin{equation}\label{q6135}
  \begin{array}{c}
    \{ B^i_{ab},B^j_{cd} \} = (1-\delta^{ij})
    \frac{1}{\upsilon_i - \upsilon_j - \varrho_{ii} + \varrho_{jj}}(B^i_{ad}B^j_{cb} - B^i_{cb}B^j_{ad})+\\ \ \\
  +\delta^{ij}\left(B^i_{ad}\delta_{bc}-B^i_{cb}\delta_{ad} + \sum_{n
    \neq i}
  \frac{B^i_{ad}B^{n}_{cb}-B^i_{cb}B^{n}_{ad}}{\upsilon_{n}-\upsilon_i - \varrho_{nn}
    + \varrho_{ii}}\right)
    \end{array}
  \end{equation}
with
 \begin{equation}\label{q6153}
  \begin{array}{c}
 x_i=\varrho_{ii} - \upsilon_i\,.
    \end{array}
  \end{equation}
The variables $\varrho_{ii}$ and $\upsilon_i$ are the Casimir functions in the AHH dual Gaudin model:
 \begin{equation}\label{eq:2}
  \begin{array}{c}
  \{\upsilon_i, A_{kl}^b \}_D = \{\upsilon_i, A_{kl}^b \} = 0 \quad
  \hbox{for } b = 1 \ldots k\,,\\
  \{ \varrho_{ii}, A_{kl}^b \}_D = \{\varrho_{ii}, A_{kl}^b \} = 0 \quad
  \hbox{for } b = 1 \ldots k\,.
    \end{array}
  \end{equation}
\end{theor}
The proof follows directly from the Dirac brackets (\ref{q3433}). Notice that $\varrho_{ii}$ are the Casimir functions.
Therefore, one can put $\varrho_{ii}=0$ and reproduce (\ref{q6214}). Then, comparing (\ref{q6152}) and (\ref{q67111}), one gets
the exact equivalence of the models:
 \begin{equation}\label{q12345}
 \begin{array}{|c|}
  \hline\\
\hbox{AHH}\Big\{zL^{\hbox{\tiny{Gaudin}}}(z)\Big\}(x)=T^{\hbox{\tiny{Heisen}}}(x){\prod\limits_j\frac{1}{x-x_j}}\,.\\ \ \\
\hline
  \end{array}
 \end{equation}

\section{Quantum Duality} \label{sec:quantum}

\subsection{Simplest example}
Let us start again  with the simplest case (${\rm sl}_2$ 4-points reduced Gaudin model and 2-cite chain) corresponding to
the four-dimensional Yang-Mills theory with $SU(2)$ gauge group. The Baxter equation for the XXX spin chain is known to
lead to the Bohr-Sommerfeld integrals that describe the SW system giving rise to the right Nekrasov prepotential \cite{Zen}.
This Baxter equation can be thought of as a quantization of the corresponding
spectral curve (\ref{q1702}):
$$
\tr\, T(x)=\tilde{y}^2z^2+H_2 \tilde{y}z+H_1=\frac{z}{1+q}K^+(\tilde{y}z)+ \frac{1}{z}\frac{q}{1+q}K^-(\tilde{y}z),
$$
where
$$
K^+(x)=(x-m_1^+)(x-m_2^+),\ \ \ K^-(x)=(x-m_1^-)(x-m_2^-).
$$
The ``quantization'' implies that (for $x=z\tilde{y}$)
\begin{equation}\label{q60}
  \tilde{y}\longrightarrow \hbar\partial_z %\frac{\p}{\p z}\equiv
\end{equation}
At this stage we need to fix the ordering. We use the normal ordering putting $\partial_z$ to the right of all $z$'s. This
provides the following rules
\begin{equation}\label{q61}
  \begin{array}{c}
    :K^+_\hbar(x): = :K^+(x):+\hbar x,
    \\
    :K^-_\hbar(x):=:K^-(x):+\hbar x,
    \\
    :\tr T_\hbar(x):=:\tr T(x):+\hbar x.
  \end{array}
\end{equation}
This recipe follows from
\begin{equation}\label{q62}
  x^2\stackrel{{\hbox{\tiny{Qaunt}}}}{\longrightarrow}
  \hbar^2 z\partial_z z\partial_z=\hbar^2z^2\partial_z^2+\hbar^2z\partial_z=
  :\!\hat{x}^2\!:+\hbar:\!\hat{x}\!:.
\end{equation}
After substituting (\ref{q60}), one has
\begin{equation}\label{q63}
  \left[\hbar^2\partial_z^2-b_\hbar(z)\hbar\partial_z+c_\hbar(z)\right]\psi_{\hbox{\tiny{Heisen}}}(z)=0,
\end{equation}
with
\begin{equation}\label{q64}
  \begin{array}{c}
    b_\hbar(z)=b(z)-\frac{\hbar}{z},\ \
    c_\hbar(z)=c(z),
  \end{array}
\end{equation}
where $b(z)$ and $c(z)$ are given by (\ref{q1704}):
 $$
 \begin{array}{l}
   b(z)=\frac{(q+1)H_2+q(m_1^++m_2^+)+(m_1^-+m_2^-)}{(q-1)(z-q)} +\frac{(m_1^-+m_2^-)}{z}-
   \frac{(q+1)H_2+(m_1^++m_2^+)+q(m_1^-+m_2^-)}{(q-1)(z-1)},
   \\
   c(z)=\frac{q+1}{q}\frac {m^-_{{1}}m^-_{{2}}-H_{{1}}}{z}+\frac{m_1^-m_2^- + qm_1^+m_2^+ -(1+q)H_1}{q(q-1)(z-q)}-
   \frac{qm_1^-m_2^- + m_1^+m_2^+ -(1+q)H_1}{(q-1)(z-1)}+\frac{m_1^-m_2^-}{z^2}.
 \end{array}
 $$
 The ``classical'' shift (\ref{q4002}) $\tilde{y}\rightarrow
 \tilde{y}+\frac{1}{2}b(z)$ is replaced here by the change
 \begin{equation}\label{q65}
   \psi_{\hbox{\tiny{Heisen}}}(z)= \psi_{\hbox{\tiny{Gaudin}}}(z)\exp\left(\frac{1}{2\hbar}\int\limits^z\!\hbox{d}z\ b_\hbar(z)\right)
 \end{equation}
 which leads to
 \begin{equation}\label{q6501}
   \left[\hbar^2\partial_z^2+c_\hbar(z)-\frac{1}{4}b^2_\hbar(z)+
   \frac{1}{2}(\partial_z b_\hbar(z))\right]\psi_{\hbox{\tiny{Gaudin}}}(z)=0.
 \end{equation}
 Making the same calculations as in the classical case, one finally gets
 \begin{equation}\label{q66}
   \left[-\hbar^2\partial_z^2+V_\hbar(z)-\frac{1-q}{z(z-1)(z-q)}H_\hbar\right]\psi_{\hbox{\tiny{Heisen}}}(z)=0,
 \end{equation}
 where $H_\hbar=H+\hbar m_q$ and
 \begin{equation}\label{q67}
   \begin{array}{l}
     V_\hbar(z)=\frac{m_0^2-\frac{\hbar^2}{4}}{z^2}+\frac{m_1(m_1-\hbar)}{(z-1)^2}+\frac{m_q(m_q-\hbar)}{(z-q)^2}-
     \frac{m_0^2+m_1(m_1-\hbar)+m_q(m_q-\hbar)-m_\infty^2}{z(z-1)}
   \end{array}
 \end{equation}
$m_0$, $m_1$, $m_q$ and $m_\infty$ being given by (\ref{q1707}). Equation (\ref{q66}) is exactly the one derived from
consideration of the conformal blocks\footnote{Different conventions on what to call the mass parameters in the gauge theory
after the $\epsilon$-deformation exist. To compare our expressions with \cite{AGT} one needs to perform the shift:
$m^{\pm}_{\mathrm{our}} = m_{\mathrm{AGT}}^{\pm} \pm \hbar/2$.}. In~\cite{MiTa} it was shown that this equation gives the quantum
periods (monodromies around the $A$ and $B$ cycles) from which one indeed obtains the correct Nekrasov functions. The two wave
functions $\psi_{\hbox{\tiny{Heisen}}}$ and $\psi_{\hbox{\tiny{Gaudin}}}$ give slightly different periods and thus different
prepotential because of the factor $b_{\hbar}$ in~(\ref{q65}). The difference for the period integrals is
 \begin{equation}
   \label{eq:802}
   \Delta\Pi = \oint b_{\hbar}(z)\,dz\,.
 \end{equation}
However, it does not affect the physics for the following reasons. Indeed, as we briefly mentioned in Section
 ~\ref{sec:reduced-phase-space} the linear Hamiltonian $H_2 =
 \sum_k\phi_k$ which enters $b_{\hbar}$ is set to vanish for the
 $\mathrm{SU}(N)$ gauge theory. The remaining terms do not depend on
 the Hamiltonians $H_i$ or equivalently on $\phi_k$ which in the gauge
 theory correspond to the vacuum moduli. Thus, the contribution $\Delta
 \Pi$ of $b_{\hbar}$ to the monodromies is a constant (depending only on
 the masses $m_k^{\pm}$), its contribution to the prepotential is
 linear in vacuum moduli and does not enter the low-energy
 effective lagrangian $\mathcal{L} = \int d^4\theta\, \mathcal{F}(\Psi)$
 due to the identity $\int d^4\theta \, \Psi = 0$.

Formulae (\ref{q65}) and (\ref{q6501}) present the relation for the ${\rm sl}_2$ Gaudin model while the ${\rm gl}_2$ case is much
simpler:
\begin{equation}\label{q6503}
   \psi_{\hbox{\tiny{Heisen}}}(z)= \psi_{\hbox{\tiny{Gaudin}}}(z)\,.
 \end{equation}
% (CITE Krichever Dhoker Phong, etc.).
%
\subsection{General case} \label{sec:general-quantum-case}
Quantization of the XXX chain spectral curve (\ref{q40}) with the SW differential (\ref{q3011}) means that $x$ should be
simply replaced by $\hbar w\partial_w$. For example, in the ${\rm GL}_2$ case one gets the Baxter equation:
 \begin{equation}
  \label{qq912}
  \left( \tr \, T(\hbar w \partial_w) - \frac{w}{1+q} {K}_{+}(\hbar
  w \partial_w) - \frac{q}{(1+q)w} {K}_{-}(\hbar
  w \partial_w) \right)\Psi^{\hbox{\tiny{Heisen}}}(w)=0\,.
 \end{equation}
Equivalently, for the Gaudin spectral curve (\ref{q3501}) the quantization is given by the replacement
$y\rightarrow\hbar\partial_z$:
 \begin{equation}\label{qq914}
  \begin{array}{l}
\left(\prod\limits_{i=1}^N(z\hbar\partial_z+\upsilon_i)+
\sum\limits_{k=1}^N\frac{\eta^1_k\xi^1_k+q\eta^q_k\xi^q_k}{q+1}\prod\limits_{i\neq
k}^N(z\hbar\partial_z+\upsilon_i) -\right.\\
 \left.-\frac{z}{q+1}\prod\limits_{i=1}^N(z\hbar\partial_z+\upsilon_i)

-z^{-1}\frac{q}{q+1}\prod\limits_{i=1}^N(z\hbar\partial_z-\mu_i)\right)\Psi^{\hbox{\tiny{Gaudin}}}(z)=0\,.
  \end{array}
 \end{equation}
Obviously, the differential operators in the brackets of (\ref{qq912}) and (\ref{qq914}) can be identified in the same way as
the classical spectral curves did.

The Baxter equation~(\ref{qq912}) looks similar to the classical equation for the spectral curve except for the substitution
$x \to \hbar z \partial_z$. All the differentials $z\partial_z$ are placed to the right of all the functions of $z$ (which is
consistent with our previous normal ordering under the replace $z\to\log z$).

To obtain a similar expression for the quantum Gaudin system one also needs to place all the differentials $z\partial_z$ to the
right. After doing this, the equivalence is evidently analogous to the classical case. Thus, the definition of the quantum
``determinant'' must be the following:
\begin{equation}
  \label{eq:9}
  \mathrm{``}\det \mathrm{"} (\hbar z \partial_z + z L_{\hbox{\tiny{Gaudin}}}(z))
  \stackrel{\mathrm{def}}{=} \mbox{all $z\partial_z$ to the right.}
\end{equation}
 \begin{theor}
The N-site  ${\rm GL}_k$ Heisenberg XXX chain  and  the
  ${\rm gl}_N$ $k+2$-point Gaudin model
are spectrally dual at the quantum level with the following relation between the wave functions:
\begin{equation}
  \label{eq:107}
  \psi_{\hbox{\tiny{XXX}}}(z) = \psi_{\hbox{\tiny{Gaudin}}}(z)\,,
 \end{equation}
i.e. the Baxter equation of the Heisenberg chain can be rewritten as the quantization of the Gaudin model spectral curve with
the ordering (\ref{eq:9}).
 \end{theor}

%Remark: In fact, this theorem follows directly from the classical equivalence of the models (\ref{q12345}) since the
%quantization of the Gaudin model spectral curve is performed in exactly the same way as the one of the Heisenberg chain. From
%this point of view, one can treat the statement (\ref{eq:107}) as a definition of the Baxter equation for the Gaudin model. On
%the other hand, one can obtain the same result following the recipe of the quantization (\ref{eq:9}).

A small problem arises only for the ${\rm sl}_N$ Gaudin model. In this case, one has an additional factor, just as in (\ref{q65}).
However, this factor is controllable. Let us compute it in the case of ${\rm GL}_2$ chain, i.e. in
the case of 4-point Gaudin model duality.

The Baxter equation for the XXX spin chain can be rewritten as a
polynomial in $\partial_z$ (instead of $z\partial_z$). To do so, one
needs to make the ordering in $(z\partial_z)^n$. To this
end, let us represent the action of $(z\partial_z)^n$ as a part of the dilatation
operator
\begin{equation}
  \label{q82}
  (z\partial_z)^n=\left.\frac{\hbox{d}^n}{\hbox{d}\alpha^n}e^{\alpha z \partial_z}
  \right|_{\alpha=0},\ \ \ e^{\alpha z \partial_z} f(z) = f(e^{\alpha} z)\,.
\end{equation}
Then, using the Taylor expansion for $f(z+z(e^{\alpha z}-1))$, one gets

\begin{equation}
  \label{q83}
  (z\partial_z)^n = \sum_{k=1}^n
   \ti C^k_n z^k \partial^k_z,\ \ \ \ti C_n^k=\sum_{l=1}^k\frac{(-1)^{k-l}l^n}{(k-l)!\,
   l!}\equiv\sum_{l=1}^k(-1)^{k-l}\frac{l^n}{k!}\left(\begin{array}{l}\!k\!\\ \!l\!\end{array}\right)\,.
\end{equation}
The coefficients $\ti C_n^k$ have the following properties:
\begin{equation}
  \label{q84}
 \ti  C_n^1=1,\ \  \ti C^{n-1}_n=\frac{n(n-1)}{2},\ \ \ti C_n^n=1.
\end{equation}
For instance, the latter one is the well-known combinatorial identity \cite{Feller}:
$$\sum\limits_{k=0}^n (-1)^{n-k}\left(\begin{array}{l}\!n\!\\
\!k\!\end{array}\right)k^r=\left\{\begin{array}{l}0,\ \hbox{if}\ r<n,\\
n!,\ \hbox{if}\ r=n.\end{array}\right.$$
%
%We introduce the coefficients of $P$ and
%$K_{\pm}$ polynomials in the following way
Finally, for $K^+(z\partial_z)$ one has
\begin{equation}
  \label{q85}
  K^+(\hbar z\partial_z)=\sum\limits_{n=1}^N
  (-1)^n\sigma_n(\{m^+\})(\hbar z\partial_z)^n=\sum\limits_{n=1}^N\sum_{k=1}^n
  (-1)^n\sigma_n(\{m^+\})\hbar^n \ti C^k_n z^k \partial^k_z
\end{equation}
and similarly for $K^-$ , where $\sigma_n(\{m^+\})$ are elementary symmetric polynomials, and $\tr T$.
Now one can rewrite the Baxter equation in the
desired form in which all $\partial_z$ are placed to the right

\begin{equation}
\label{q86} \sum\limits_{n=1}^N\sum_{k=1}^n
  \left(H_{n+1} - \frac{z(-1)^n}{q+1}\sigma_n(\{m^+\}) -
  \frac{q(-1)^n}{(1+q) z}\sigma_n(\{m^-\})\right)\hbar^n \ti C^k_n z^k \partial^k_z\,\psi=0
\end{equation}
Analogously to the transition from (\ref{q14}) to (\ref{q15}), one may write the equation in the form:
\begin{equation}\label{q87}
\left(\hbar^N\partial_z^N-b_\hbar(z)\hbar^{N-1}\partial_z^{N-1}+\sum\limits_{m=0}^{N-2}c_m(z)\hbar^m\partial_z^m\right)\psi=0
\end{equation}
From (\ref{eq:9}), (\ref{q82}) and (\ref{q84}) it follows that
\begin{equation}
  \label{q88}
  b_{\hbar} (z) = \frac{(1+q)}{(z-1)(z-q)} \left( {H_N} +
    \frac{z \sum_{k=1}^N m_k^{+}}{1+q} + \frac{q \sum_{k=1}^{N}
      m_k^{-}}{(1+q) z } \right) - \hbar \frac{N(N-1)}{2z}\,.
\end{equation}

Again, one may eliminate $b_\hbar(z)$ from the equation. By the construction, we come to
 \begin{theor} {\bf \cite{MMZZ}}
The N-site  ${\rm GL}_2$ Heisenberg XXX chain defined by  (\ref{q1})-(\ref{q1601}) and  the
  ${\rm gl}_N$ Gaudin model
  (\ref{q25})-(\ref{q42}) are spectrally dual at the quantum level with the following relation between the wave functions:
\begin{equation}
  \label{eq:108}
  \psi_{\hbox{\tiny{XXX}}}(z) = \psi_{\hbox{\tiny{Gaudin}}}(z)
  e^{\frac{1}{N\hbar} \int^z b_{\hbar} \,dz}.\,,
 \end{equation}
i.e. the Baxter equation of the Heisenberg chain can be rewritten as the quantization of the Gaudin model spectral curve with
the ordering (\ref{eq:9}).
 \end{theor}
Indeed, all the classical statements work with our choice of ordering (\ref{eq:9}) for the Gaudin system. Thus, the two
systems are exactly equivalent.

\section{Comments and Discussion}
\label{sec:discussion}

In this section we discuss possible generalizations and relations of the models and the corresponding duality. First of all,
there is a wide class of the monodromy preserving equations which can be considered as some generalization of the Gaudin model.
\begin{itemize}

\item{\bf Painlev\'e-Schlesinger Equations.} In the ${\rm{gl}}_2$ case, the coset space (\ref{q27}) is the phase space of the Painlev\'e VI equation \cite{Painleve}. The
Painlev\'e equations can be interpreted as equations describing isomonodromic deformations \cite{Painleve2} of the linear
system
 \begin{equation}\label{q4201}
\left(\partial_z-L^G(z)\right)\Psi(z,q)=0.
 \end{equation}
The "deformation" generates dynamics with respect to the marked points. In the case of the 4-points Gaudin model, it is only $q$:
 \begin{equation}\label{q4202}
\left(\partial_q-M^G_q(z)\right)\Psi(z,q)=0,\ \ M^G_q(z)=-\frac{A^q}{z-q}.
 \end{equation}
The zero-curvature condition for the above equations is the monodromy preserving equation:
 \begin{equation}\label{q4203}
\partial_q L^G(z)-\partial_z M^G_q(z)+[L^G(z),M^G_q(z)]=0.
 \end{equation}
The residues at $0,1,q$  provide the finite-dimensional equations on $A^c$ which are called the Schlesinger equations:
 \begin{equation}\label{q42031}
   \begin{array}{l}
    \partial_q A^0=-\frac{1}{q}[A^0,A^q],\\
    \partial_q A^1=\frac{1}{1-q}[A^1,A^q],\\
    \partial_q A^q=\frac{1}{q}[A^0,A^q]-\frac{1}{1-q}[A^1,A^q].
   \end{array}
 \end{equation}
After the reduction (\ref{q27}) the later equations reduce to the single one: the Painlev\'e VI equation.
%This is why we refer to this example as the Painlev\'e VI.
All the same can be done in the general ${\rm{gl}}_N$ case with any number of marked points.
%The similar procedure leads to the higher Painlev\'e equations.
In \cite{Yamada} Y.Yamada\footnote{This type of equation appeared earlier within the context of the
  UC-hierarchy in \cite{Tsuda}.} considered the higher Painlev\'e equations corresponding to the configuration
(\ref{q29})-(\ref{q33003}). Note that if one replaces $\partial_q$ by $\partial_t$ in all the above equations,
(\ref{q42031}) describes the flow in the unreduced Gaudin model, while the reduced one is just the autonomous version of the
corresponding Painlev\'e equation(s). The Schlesinger system is non-autonomous version of the Gaudin model:
 \begin{equation}\label{q7110}
\p_{t_j}A_i=\frac{[A_i,A_j]}{z_i-z_j}\ \ \ \rightarrow\ \ \ \p_{z_j}A_i=\frac{[A_i,A_j]}{z_i-z_j}
 \end{equation}
Relations between the Gaudin and Schlesinger systems can be summarized as follows:
 \begin{equation}\label{q7111}
\left\{\begin{array}{l} (\lambda-L(z))\Psi=0\\
(\p_{t_i}-M_i(z))\Psi=0\end{array}\right.\ \ \ \stackrel{\lambda\rightarrow\p_z}{\longrightarrow}\ \ \ \
\left\{\begin{array}{l} (\p_z-L(z))\Psi=0\\
(\p_{z_i}-M_i(z))\Psi=0\end{array}\right.\,.
 \end{equation}
 \begin{equation}\label{q7112}
    \begin{array}{l}
\p_{t_i}L=[L,M_i]\ \ \ \rightarrow\ \ \ \ \p_{z_i}L-\p_z M=[L,M_i]
\\ \ \\
\{H_i,H_j\}=0\ \ \ \rightarrow\ \ \ \ \partial_{t_i}H_j-\partial_{t_j}H_i=\{H_i,H_j\}
   \end{array}
 \end{equation}
Relation between the Schlesinger system and the Painlev\'e-type equations is given by the reduction procedure:
 $$
\begin{array}{l}
\hbox{Unreduced Gaudin}\ \ \longrightarrow\ \ \hbox{Schlesinger System}
\\
\hspace{15mm}\downarrow\hspace{45mm}  \downarrow
\\
\ \ \hbox{Reduced Gaudin}\ \ \longrightarrow\ \ \hbox{Painlev\'e Equations}
\end{array}
 $$
Let us list open problems arising from the Gaudin-Schlesinger correspondence:

1. The recipe (\ref{q7111}) relates the isomonodromic deformations with the quantization problem. The corresponding {\em
"isomonodromic quantization"} was studied in \cite{iso}. The natural quantization is given by the
Knizhnik-Zamolodchikov-Bernard (KZB) equations for the Gaudin model. We hope to find a relation between solutions to the Baxter
equation and the corresponding KZB equation.
%In the same time the method of quantization of the unreduced Gaudin
%model suggested in \cite{Talalaev} also uses the rule $\lambda\rightarrow\p_z$.
%and
%$$
%\frac{du}{dt^2}=\nu\wp'(u,\tau)\ \ \ \rightarrow\ \ \ \frac{du}{d\tau^2}=\nu\wp'(u,\tau)
%$$

2. In \cite{ZZ} it was shown that the classical linear problem at the r.h.s. of (\ref{q7111}) leads to the non-stationary quantum
problem for the reduced model written in spectral variables. At the same time, V.Bazhanov and V.Mangazeev showed \cite{BaMa}
that the Baxter equation of some special chain is compatible with the Lam{\'e} equation. Therefore, one can expect that the
Baxter equations are compatible with some kind of the non-stationary Schr\"odinger equation.

3. In \cite{Yamada} the higher (coupled) Painlev{\'e}-type equations related to the specific
configuration  (\ref{q29})-(\ref{q33003}) were studied. In the simplest case the underlying autonomous system is the $BC_1$
Calogero-Inozemtsev model, while its isomonodromy version is the Painlev\'e VI equation in the elliptic form \cite{Painl}.
For the ${\rm
gl}_N$ case one can also hope to find an elliptic integrable ("spinless") system of Calogero-Gaudin type and the corresponding
Painlev{\'e}-type system with $N-1$ degrees of freedom and $2N$ coupling constants.

4. The elliptic form of the Painlev{\'e} VI equation can be presented as a non-autonomous version of the 1-site XYZ chain
interacting with a
constant external magnetic field\footnote{The corresponding mechanical system is the Zhukovsky-Volterra gyrostat.}, i.e.
with dynamical boundary conditions satisfying the reflection equation \cite{LOZ2}.
Therefore, the spectral duality provides some relation between
the XYZ and XXX models which we hope to investigate.

\item {\bf Quantization.} It should be mentioned that we do not impose any boundary conditions which provide a
valuable quantum problem, i.e. we do not specify wave functions explicitly. We compare the Baxter equations which are the
quantizations of the spectral curves written in separated variables. Alternatively, one can specify the spaces of solutions
initially and then verify their identification through the duality transformation. This is the recipe of \cite{MTV} where the
authors considered very close problem in terms of the Bethe vectors. Recall, that the Lax matrix of the n-point reduced
Gaudin model is of the form
\begin{equation}\label{q73}
\begin{array}{c}
 L^G(z)=\sum\limits_{c=1}^{n-1} \frac{A^c}{z-z_c}=\frac{1}{z}\left(-\Upsilon+\sum\limits_{c=2}^n
 \frac{z_cA^c}{z-z_c}\right)\,,
 \end{array}
 \end{equation}
where $z_1$ is set to zero and $\Upsilon$ is the residue at $\infty$. The expression in the brackets (at the r.h.s. of
(\ref{q73})) is similar to the case of Gaudin model considered by E.Mukhin, V.Tarasov and A.Varchenko. In \cite{MTV} they
conjectured a duality between the XXX chain and trigonometric Gaudin model at the quantum level, which relates the
corresponding Bethe vectors. The derivation uses the space of quasi-polynomials (and quasi-exponentials), i.e. the explicit form
of solutions to the Baxter equations. It is based on the linear (Lie algebra) commutation relations, while the Gaudin model
under consideration here is the rational (although reduced) one. The Poisson structure (which is discussed at the
classical level) is quadratic. Therefore, one can expect to find a relation between the results of this paper and those of
\cite{MTV} by use of a quantum version of the Dirac reduction. At the classical level, the main difference comes from the
definition of the SW differential. In our case $\hbox{d}S=\lambda\hbox{d}z$ with the change of variables to the XXX chain: $z=w$,
$\lambda=x/w$. In \cite{MTV} the differential (presumably) is $\hbox{d}S=\lambda\frac{\hbox{d}z}{z}$ and the corresponding
change of variables is $z=w$, $\lambda=x$. Notice also that the expression in the brackets (at the r.h.s. of (\ref{q73}))
contains the constant term ($\Upsilon$), which (without the factor $1/z$) leads to the second order pole at $\infty$ for
$L^G(z)\hbox{d}z$. This type of Gaudin models was studied in \cite{FFT}.

Besides the approach proposed here, different recipes for quantizations of the Gaudin model were suggested, e.g., in
\cite{FFR} and \cite{Talalaev}). The recipes proposed there are valid for the unreduced Gaudin models. Therefore, one needs to
perform the quantum reduction procedure in order to relate the results of \cite{FFR} and \cite{Talalaev} with ours.

\item {\bf{Higher spin chains.}} In this paper we consider the higher spin chains with orbits of the minimal rank at each site,
i.e. $S^k\in{\rm sl}_k$ is conjugated to the element of the form $\hbox{diag}(r,...,r,-(k-1)r)$. It is also interesting to
describe the spectral duality for the generic ${\rm GL}_k$ spin chain. We hope to solve this problem in future
publications.

\item {\bf Anisotropic chains.}
Another interesting generalization is induced by the five-dimen\-sional AGT \cite{5dAGT} which implies a correspondence
between the XXZ magnets (see \cite{BPMY}, \cite{SW5d}) and a Gaudin-like model with relativistic (difference) dynamics. This
latter would emerge, since on the conformal side one deals in this case with the q-Virasoro conformal block which implies a
difference Schr\"odinger equation for the block with insertion of the degenerate field.

A duality for the five dimensional quiver gauge theories was proposed in \cite{BPMY}. It relates the theories with the
gauge groups
$SU(N)^{M-1}$ and $SU(M)^{N-1}$ compactified on $\mathbb R^4 \times S^1$, the radius of $S^1$ being $R_5$. At the level of
integrable mechanics, the relevant system in this case is the XXZ spin chain. For the case of single $SU(N)$ gauge group it
can be written as follows \cite{SW5d}:
\begin{equation}
  w + Q_{2N}(e^{2\zeta/R_5}) w^{-1} = P_N(e^{2\zeta/R_5})\,,
\end{equation}
where $Q_{2N}(\lambda) = \prod_{i=1}^N (\lambda - e^{2m_i^{+}/R_{5}})(\lambda - e^{2m_i^{-}/R_5})$ and $P_N(\lambda) =
\prod_{i=1}^{N} (\lambda - e^{a/R_5}))$ is a polynomial of degree $N$ with the coefficients parameterizing the Coulomb branch
of the vacuum moduli space. One observes that the curve has exactly the same form as in the four dimensional case, except
that it is written in terms of the variables $(w,e^{2\zeta})$. However, the SW differential is different, namely $dS = \zeta
dw/w$.

From the results of \cite{BPMY} one gets the duality transformation for the spectral curve of the XXZ spin chain. It relates
two different XXZ systems, corresponding to the two sides of the $SU(N)^{M-1} \leftrightarrow SU(M)^{N-1}$ duality, that is,
the $N$-site $\rm{gl}_M$ spin chain and the $M$-site $\rm{gl}_N$ spin chain with the spin matrices of the minimal rank. More
concretely, the duality exchanges the variables $w \leftrightarrow e^{2\zeta/R_5}$. The SW differential is manifestly
invariant (up to a sign) under this transformation: $dS = \ln e^{2\zeta/R_5} d \ln w \cong - d\ln e^{2\zeta/R_5} \ln w$.

We are going to describe this duality explicitly in our future publication \cite{MMRZZ2}. An extension to six dimensions
(an elliptic extension of the differential operator in the Schr\"odinger equation versus the XYZ magnet) is also extremely
interesting to construct.

\end{itemize}

%\section*{Appendix A: Classical Spectral Curve}\label{App_A}
%\addcontentsline{toc}{section}{Appendix A: Classical Spectral Curve}
%\def\theequation{A\arabic{equation}}
%\setcounter{equation}{0}

}

\begin{thebibliography}{99}
\addcontentsline{toc}{section}{References}

\footnotesize{

\bibitem{MMZZ} A.Mironov, A.Morozov, Y.Zenkevich and A.Zotov, arXiv:1204.0913

\bibitem{AGT}
D.Gaiotto, %{\em N=2 dualities},
arXiv:0904.2715.

L.F.Alday, D.Gaiotto and Y.Tachikawa, Lett.Math.Phys. { 91} (2010) 167-197, arXiv:0906.3219.

\bibitem{SWint} A.Gorsky, I.Krichever, A.Marshakov, A.Mironov and A.Morozov,
Phys.Lett., { B355} (1995) 466-477, hep-th/9505035;\\
R.Donagi and E.Witten, Nucl.Phys., { B460} (1996) 299-334, hep-th/9510101.

\bibitem{NS} N.Nekrasov and S.Shatashvili,  arXiv:0908.4052;\\
N.Nekrasov, A.Rosly and S.Shatashvili,  Nucl.Phys. (Suppl.) { B216}} (2011) 69-93, arXiv:1103.3919.

\bibitem{BS} A.Mironov and A.Morozov, %Bohr-Sommerfeld,
%\emph{Nekrasov Functions and Exact Bohr-Zommerfeld Integrals},
JHEP { 04} (2010) 040, arXiv:0910.5670;
%\emph{Nekrasov Functions from Exact BS Periods: the Case of SU(N)},
J.Phys. { A43} (2010) 195401, arXiv:0911.2396.

\bibitem{AGTN} N.Wyllard, JHEP 11 (2009) 002, arXiv:0907.2189\\
A.Mironov and A.Morozov, Phys.Lett. { B680} (2009) 188-194, arXiv:0908.2190; Nucl.Phys. B825 (2009) 1-37, arXiv:0908.2569.

\bibitem{AGTproof} A.Mironov, A.Morozov and Sh.Shakirov, JHEP {\bf 1102} (2011) 067, arXiv:1012.3137;\\
A.Belavin, V.Belavin, Nucl.Phys.B850:199-213,2011, arXiv:1105.5800

\bibitem{SO} L.Alday, D.Gaiotto, S.Gukov, Y.Tachikawa and H.Verlinde, JHEP {
1001} (2010) 113,\\
arXiv:0909.0945;\\
A.Braverman, arXiv:math/0401409;\\
A.Braverman and P.Etingof, arXiv:math/0409441;\\
V.Fateev and I.Litvinov, JHEP { 1002} (2010) 014, arXiv:0912.0504;\\
C.Kozcaz, S.Pasquetti and N.Wyllard, arXiv:1004.2025.

\bibitem{AGTmamo} R.Dijkgraaf and C.Vafa, arXiv:0909.2453;\\
H.Itoyama, K.Maruyoshi and T.Oota,
%\emph{Notes on the Quiver Matrix Model and 2d-4d Conformal Connection},
Prog.Theor.Phys. { 123} (2010) 957-987, arXiv:0911.4244;\\
T.Eguchi and K.Maruyoshi,
%\emph{Penner Type Matrix Model and Seiberg-Witten Theory},
arXiv:0911.4797;
%{\it Seiberg-Witten theory, matrix model and AGT relation},
arXiv:1006.0828;\\
R.Schiappa and N.Wyllard,
%\emph{An $A_r$ threesome: Matrix models, $2d$ CFTs and $4d$ N=2 gauge
%theories},
arXiv:0911.5337;\\
A.Mironov, A.Morozov and Sh.Shakirov,
%\emph{Matrix Model Conjecture for Exact BS Periods and Nekrasov Functions},
JHEP { 02} (2010) 030, arXiv:0911.5721;
%\emph{Conformal blocks as Dotsenko-Fateev Integral Discriminants},
Int.J.Mod.Phys. { A25} (2010) 3173-3207, arXiv:1001.0563.

\bibitem{BSmore} A.Popolitov, arXiv:1001.1407;\\
Wei He and Yan-Gang Miao,
%{\it Magnetic expansion of Nekrasov theory: the SU(2) pure gauge theory},
Phys.Rev. { D82} (2010) 025020, arXiv:1006.1214;\\
F.Fucito, J.F.Morales, R.Poghossian and D. Ricci Pacifici,
%{\it Gauge theories on $\Omega$-backgrounds from non commutative
%Seiberg-Witten curves},
arXiv:1103.4495;\\
N.Dorey, T.J.Hollowood and S.Lee, arXiv:1103.5726;\\
M.Aganagic, M.Cheng, R.Dijkgraaf, D.Krefl and C.Vafa, arXiv:1105.0630.

\bibitem{Zen} Y.Zenkevich, Phys.Lett.B 701:630-639 (2011), arXiv:1103.4843

\bibitem{MiTa} K.Maruyoshi and M.Taki, Nucl.Phys. B841 (2010) 388-425,
arXiv:1006.4505.

\bibitem{Yamada} Y.Yamada, J.Phys. A44 (2011) 055403, arXiv:1011.0292.

\bibitem{5} A.Marshakov, A.Mironov and A.Morozov,  J.Geom.Phys. 61 (2011)
1203-1222, arXiv:1011.4491.

\bibitem{Mironov:2010qe} A.Mironov, A.Morozov and S.Shakirov, Int.J.Mod.Phys.
A27 (2012) 1230001, arXiv:1011.5629.

\bibitem{Tai} K.Muneyuki, Ta-Sheng Tai, N.Yonezawa and R.Yoshioka,  JHEP 2011
(2011) 125, arXiv:1107.3756.

\bibitem{Tai2} Ta-Sheng Tai,  JHEP 10 (2010) 107, arXiv:1008.4332.

\bibitem{BPMY} L.Bao, E.Pomoni, M.Taki, F.Yagi, arXiv:1112.5228

\bibitem{BBFLT} A.Belavin, M.Bershtein, B.Feigin, A.Litvinov, G.Tarnopolsky, arXiv:1111.2803

\bibitem{Min-xin Huang} Min-xin Huang, arXiv:1205.3652

\bibitem{Bourgine} Jean-Emile Bourgine, arXiv:1206.1696

\bibitem{mty}
Kohei Motegi, Ta-Sheng Tai, Reiji Yoshioka,
%On non-stationary Lam\'e equation from WZW model and spin-1/2 XYZ chain. . Feb 2012. 16 pp. e-Print:
arXiv:1202.1764 [hep-th]

\bibitem{LMNS} G.Moore, N.Nekrasov and S.Shatashvili, Nucl.Phys. { B534}
(1998)
549-611, hep-th/9711108; hep-th/9801061;\\
A.Losev, N.Nekrasov and S.Shatashvili, Commun.Math.Phys. { 209} (2000) 97-121, hep-th/9712241;  hep-th/9803265.

\bibitem{Nekr} N.Nekrasov, Adv.Theor.Math.Phys. { 7} (2004) 831-864,
hep-th/0206161;\\
N.Nekrasov and A.Okounkov, hep-th/0306238.

\bibitem{SWsc} A.Gorsky, A.Marshakov, A.Mironov and A.Morozov, Phys.Lett. {
B380} (1996) 75-80, arXiv:hep-th/9603140; arXiv:hep-th/9604078.

\bibitem{Heisen} W.Heisenberg, Zeitschrift f\"ur Physik, 49 (9-10) (1928) 619.
%H.Bethe, Zeitschrift f\"ur Physik, 76, 205 (1931)

\bibitem{SW} N.Seiberg, E.Witten, Nucl.Phys., B426 (1994) 19-52,
hep-th/9407087;\\
%
N.Seiberg, E.Witten, Nucl.Phys., B431 (1994) 484-550, hep-th/9408099.

\bibitem{Gaudin1} R.Garnier,
Rend. del Circ. Matematice Di Palermo, 43, Vol. 4 (1919);\\
%
M.Gaudin, Jour. Physique, 37 (1976) 1087-1098.

\bibitem{AHH}
M.R.Adams, J.Harnad and J.Hurtubise, Lett.Math.Phys. 20 (1990) 299-308
(1990);\\
%
J.Harnad, Comm.Math.Phys. 166 (1994) 337-365, hep-th/9301076.

\bibitem{Harnad2} M.Bertola, B.Eynard and J.Harnad, Comm.Math.Phys. 229 (2002)
73-120, nlin/0108049.

\bibitem{FT} L.Faddeev, L.Takhtajan, {\em Hamiltonian approach to solitons theory}, Nauka, Moscow, 1986 (in
Russian) and Berlin: Springer-Verlag. 1987.

\bibitem{W1} G.Wilson,  J. Reine Angew. Math. 442 (1993) 177-204.


\bibitem{Dirac}
P.A.M. Dirac, Proc. Roy. Soc. London, ser. A, 246, 326 (1950)\\
M.Henneaux and C.Teitelboim, {\em Quantization of Gauge Systems}, Princeton University Press (1994).

\bibitem{BDOZ}
H.W.Braden, V.A.Dolgushev, M.A.Olshanetsky and A.V.Zotov,
%{\em Classical r-matrices and the Feigin–Odesskii algebra via Hamiltonian and Poisson reductions},
J. Phys. A: Math. Gen. 36 6979-7000 (2003), hep-th/0301121

L.Feh{\'e}r, A.G{\'a}bor and B.G.Pusztai, J. Phys. A: Math. Gen. 34 7235 (2001),     arXiv:math-ph/0105047


\bibitem{Hit} N.J.Hitchin, G.B.Segal, R.S.Ward, {\em Integrable systems: Twistors, loop groups, and Riemann surfaces},
Clarendon Press, Oxford (1999).

\bibitem{Ar} V.Arnold,
{\em Mathematical Methods in Classical Mechanics}, Springer, 1978.

\bibitem{Skl4} E.Sklyanin,
%{\em Some algebraic structures connected with the Yang—Baxter equation}
Functional Analysis and Its Applications Vol. 16, Num. 4, 263-270 (1982).


\bibitem{STS} M.A.Semenov-Tyan-Shanskii, Functional Analysis and Its Applications
Vol. 17, Num. 4 (1983), 259-272
% What is a classical r-matrix?


\bibitem{KRS} P.P.Kulish, N.Yu.Reshetikhin and E.K.Sklyanin, Lett.Math.Phys. Vol. 5, Num. 5 (1981), 393-403.
%Yang-Baxter equation and representation theory: I


\bibitem{Krich1} I.Krichever,
%{\em Methods of algebraic geometry in the theory of non-linear equations},\\
Russ.Math.Surv., 32, 185 (1977)

B.A.Dubrovin, I.M.Krichever, S.P.Novikov,
%{\em Integrable systems. I, Current Problems in Mathematics. Fundamental Directions},
Vol. 4, Itogi Nauki i Tekhniki, Akad. Nauk SSSR, Vsesoyuz. Inst. Nauchn. i Tekhn. Inform., Moscow, 1985,
179–-284 (in Russian).

\bibitem{Krich2} I.M. Krichever and D.H.Phong,     J.Diff.Geom. 45 (1997) 349-389,     arXiv:hep-th/9604199\\
%On the integrable geometry of soliton equations and N=2 supersymmetric gauge theories
I.M. Krichever and D.H.Phong,  arXiv:hep-th/9708170.

\bibitem{SoV} E.K.Sklyanin, arXiv:solv-int/9504001

\bibitem{SoV2} H.Flaschka, D.W.McLaughlin, Progr.Theor.Phys., 55, 438-456 (1976)

I.M.Gel'fand, L.A.Diki, Funct.Anal.Appl., 13,  8-20 (1979)

S.P.Novikov, A.P.Veselov, Proc.Steklov Inst.Math., 3, 53-65 (1985)

O.Babelon, {\em A short introduction to classical and quantum integrable systems}, (2007)

A.V.Tsiganov, {\em Separation of variables in integrble systems}, Izhevsk (2005)


\bibitem{LOZ1} A.Levin, M.Olshanetsky, A.Zotov,
%{\em Hitchin Systems - Symplectic Hecke Correspondence and Two-dimensional Version},
Comm.Math.Phys. 236 (2003) 93-143, arXiv:nlin/0110045

A.Levin, M.Olshanetsky, A.Smirnov, A.Zotov,
%{\em Characteristic Classes and Integrable Systems. General Construction},
arXiv:1006.0702; arXiv:1007.4127


\bibitem{APS} P.Argyres, M.Plesser and A.Shapere, Phys. Rev. Lett. 75, 1699 (1995) arXiv:hepth/9505100.

\bibitem{GGM} A.Gorsky, S.Gukov and A.Mironov,
Nucl. Phys. B517 (1998) 409-461, arXiv:hep-th/9707120;\\
%
A.Gorsky and A.Mironov, hep-th/0011197.
%%CITATION = HEP-TH/0011197;%%



%\bibitem{Nekr} N.Nekrasov, {\em Seiberg-Witten Prepotential From Instanton
%Counting}, Adv. Theor. Math. Phys. 7, 831?-864 (2004); arXiv:hep-th/0206161

\bibitem{Baxter} R.J. Baxter, {\em Exactly Solved Models in Statistical
Mechanics}, Academic Press, London (1982).


\bibitem{Skl2}
E.Sklyanin, L.Takhtajan, L.Faddeev, Theor. Math. Phys. 40 (1979) 194;\\
E.Sklyanin,
%{\em Quantum inverse scattering method. Selected topics},
arXiv:hep-th/9211111
\\
P.Kulish, E.Sklyanin,
%{\em Quantum inverse scattering method and the Heisenberg ferromagnet},
Phys. Lett. A, Vol. 70, Iss.5–6, 461–463 (1979)
\\
A.Izergin, V.Korepin,
%{\em The inverse scattering method approach to the quantum Shabat-Mikhailov model},
Comm.Math.Phys. Vol. 79, Num. 3, 303-316 (1981)
\\
L.D. Faddeev, Lectures on quantum inverse scattering method, World Scientific, Singapore, 1990


\bibitem{FFR} B.Feigin, E.Frenkel and N.Reshetikhin,  Comm.Math.Phys. 166
(1994) 27-62.

\bibitem{Talalaev} D.Talalaev, Func.Anal.Appl. 40 (2006) 73-77 (2006);\\
%
A.Chervov and D.Talalaev, hep-th/0604128.


\bibitem{Grun} F.A.Gr\"{u}nbaum,
%{\em The limited angle reconstruction problem in computed tomography},
Proc.Symp.Appl.Math., Vol. 27, AMS, L. Shepp (ed.), pp. 43-61 (1982)
\\
F.A.Gr\"{u}nbaum,
%{\em A new property of reproducing kernels for classical orthogonal polynomials},
J. Math. Anal. Appl. 95, 491-500 (1983)


\bibitem{Duis} J.J.Duistermaat, F.A.Gr\"{u}nbaum,
%{\em Differential equations in the spectral parameter},
Comm.Math.Phys., Vol. 103, Num. 2, 177-240 (1986)

%\bibitem{W1} G.Wilson, {\em Bispectral commutative ordinary differential operators}, J. Reine Angew.
%Math. 442, 177–204 (1993)

\bibitem{AMM} H.Airault, H.P.McKean, J.Moser,
%{\em Rational and elliptic solutions of the Korteweg–de Vries equation and a related many-body problem},
Comm.Pure Appl.Math. 30, 95–148 (1977)

\bibitem{Wilson} G.Wilson,
%{\em Collisions of Calogero-Moser particles and an adelic Grassmannian (With an Appendix by I.G. Macdonald)},
Invent. math. 133, 1-41 (1998)

\bibitem{Kasman} A.Kasman,
%{\em Bispectral KP solutions and linearization of Calogero–Moser particle systems},
Commun.Math.Phys. 172, 427–448 (1995),     arXiv:hep-th/9412124


\bibitem{bs0} B.Bakalov, E.Horozov, M.Yakimov,
%{\em General methods for constructing bispectral operators},
Phys.Lett.A, Vol. 222, Issues 1–2, 21, 59–66 (1996), arXiv:q-alg/9605011

V.M.Buchstaber, V.Z.Enolskii, D.V.Leykin,
%{\em Rational analogs of abelian functions},
Functional Analysis and Its Applications, Vol. 33, Num. 2, 83-94 (1999)

J.P.Zubelli, F.Magri,
%{\em Differential equations in the spectral parameter, Darboux transformations and a hierarchy of master symmetries for KdV },
Comm.Math.Phys., Vol. 141, Num. 2, 329-351 (1991)

V.Spiridonov, A.Zhedanov,
%{\em Discrete Darboux transformations, the discrete-time Toda lattice, and the Askey-Wilson polynomials},
Methods and Applications of Analysis 2 (4),  369–398 (1995)

B.Bakalov, E.Horozov, M.Yakimov,
%{\em Bispectral Algebras of Commuting Ordinary Differential Operators },
Comm.Math.Phys., Vol. 190, 331-373 (1997), q-alg/9602011

{\em The bispectral problem}, J.Harnad, A.Kasman (eds.), CRM Proc. Lecture Notes 14, Amer. Math. Soc., Providence, RI, (1998)


\bibitem{MM5} A.Mironov and A.Morozov,        Phys.Lett. B475 (2000) 71-76,     arXiv:hep-th/9912088;
 arXiv:hep-th/0001168;   Phys.Lett. B524 (2002) 217-226,  arXiv:hep-th/0107114



\bibitem{p-q} S.N.M.Ruijsenaars,
%{\em Action-Angle Maps and Scattering Theory for Some Finite-Dimen\-sio\-nal Integrable Systems},
Comm.Math.Phys. 115, 127-165 (1988)

H.W.Braden, A.Marshakov, A.Mironov, A. Morozov,
%{\em On double-elliptic integrable systems: 1. A duality argument for the case of SU(2)},
Nucl.Phys. B, Vol. 573, Issues 1–2, 1, 553–-572 (2000),     arXiv:hep-th/9906240

V.Fock, A.Gorsky, N.Nekrasov, V.Rubtsov,
%{\em Duality in integrable systems and gauge theories},
JHEP07(2000)028, arXiv:hep-th/9906235

L Feh{\'e}r and C Klimcik, J. Phys. A: Math. Theor. 42 185202 (2009),     arXiv:0901.1983 [math-ph]




\bibitem{Gaudin2} M.Gaudin, Jour. Physique, 37, 1087-1098 (1976)


\bibitem{KrB}
M.Jimbo, T.Miwa, Y.Mori, M.Sato,
%{\em Density Matrix of an Impenetrable Bose Gas and the Fifth Painlev+e Transcendent},
Physica 1D, 80-158 (1980)

I.Krichever, O.Babelon, E.Billey, M.Talon,
%{\em Spin generalization of the Calogero-Moser system and the Matrix KP equation},
AMS Transl. (2), Vol. 221, 83-120 (1995); arXiv:hep-th/9411160

N.Nekrasov,
%{\em Infinite-dimensional algebras, many-body systems and gauge theories},
AMS Transl. (2), Vol. 191 (1999)


\bibitem{Painl} P.Painlev\'e, C.R.Acad.Sci. (Paris) { 143} (1906) 1111-1117;

Yu.Manin, AMS Transl. (2), { 186} (1998) 131-151;
%

V.Inozemtsev, Lett.Math.Phys. { 17} (1989) 11-17;
%

A.Zotov, Lett.Math.Phys. { 67} (2004) 153-165, hep-th/0310260;
%

A.Levin and A.Zotov, AMS Transl. (2), 221, (2007) 173-184;


\bibitem{ZZ} A.Zabrodin and A.Zotov, arXiv:1107.5672.


\bibitem{Feller} W.Feller, {\em An introduction to the theory of probability and its applications}, Wiley, New York, 1967
%Mir, Moscow, 1967



\bibitem{LOZ2} A.Levin, M.Olshanetsky and A.Zotov, Commun.Math.Phys. 268 (2006) 67-103,
math.QA/0508058;\\
Yu.Chernyakov, A.Levin, M.Olshanetsky and A.Zotov,
%{\em Elliptic Schlesinger system and Painlev\'e VI}
J. Phys. A: Math. Gen. 39 12083 (2006),     arXiv:nlin/0602043

%\bibitem{MMRZZ} A.Mironov, A.Morozov, B.Runov, Y.Zenkevich and A.Zotov,
%{\em Spectral Duality Between Heisenberg Chain and Gaudin Model}, to appear.


\bibitem{Painleve}
P.Painlev\'e, Acta Math., { 21} (1902) 1-85
\\
R.Fuchs, C. R. Acad. Sci. (Paris) { 141} (1905) 555-558
\\
B.Gambier, C. R. Acad. Sci. (Paris), {142}, 266-269, (1906)

\bibitem{Painleve2}
L.Schlesinger,  J. Reine u. Angew. Math. ,{141}, 96-145 (1912)
\\
M.Jimbo and T.Miwa,  Physica D, {2}, 407-448  (1981)



\bibitem{Tsuda} T.Tsuda, Comm.Math.Phys. 248 (2004) 501-526; arXiv:1007.3450.

\bibitem{iso} N.Reshethikin, Lett. Math. Phys. 26 (1992) 167
\\
M. Jimbo, T.Miwa and K.Ueno, Physica 2D (1981) 306
\\
M. Jimbo, T.Miwa, Y.Mˆori and M. Sato, Physica 1D (1980) 80
\\
J.Harnad, preprint CRM-2890 (1994), hep-th/9406078.
\\
D.A.Korotkin, J.A.H.Samtleben,     arXiv:hep-th/9511087
\\
J.Teschner,     arXiv:1005.2846

\bibitem{MTV} E.Mukhin, V.Tarasov, A.Varchenko,
math.QA/0510364; Advances in Mathematics, 218 (2008) 216-265, math.QA/0605172.

\bibitem{FFT} B.Feigin, E.Frenkel, V.Toledano-Laredo, Adv.Math.223:873-948 (2010), math.QA/0612798

\bibitem{5dAGT} H.Awata and Y.Yamada, JHEP { 1001} (2010) 125,
arXiv:0910.4431; arXiv:1004.5122;\\
S.Yanagida, arXiv:1005.0216;\\
A.Mironov, A.Morozov, Sh.Shakirov, A.Smirnov, Nucl.Phys. { B855} (2012) 128-151,     arXiv:1105.0948

\bibitem{SW5d} A.Gorsky, S.Gukov and A.Mironov,
%Supersymmetric Yang-Mills
%theories, integrable systems and their stringy/brane origin-II,
%hep-th 9710239 to appear in Nucl.Phys.B
Nucl.Phys., { B518} (1998) 689, arXiv:hep-th/9710239;\\
A.Marshakov and A.Mironov,
%Prepotentials in 5d and 6d theories from integrable systems, hep-th 9711239
Nucl.Phys., { B518} (1998) 59-91, hep-th/9711156.

\bibitem{BaMa} V.Bazhanov and V.Mangazeev, J. Phys. A: Math. Gen. 38 L145 (2005); arXiv:hep-th/0411094


\bibitem{MMRZZ2} A.Mironov, A.Morozov, B.Runov, Y.Zenkevich and A.Zotov,
{\em Spectral Duality Between XXZ Chains and 5d Gauge Theories}, to appear

































\end{thebibliography}
\end{document}